\documentclass{article}

\usepackage{arxiv}

\usepackage[utf8]{inputenc} % allow utf-8 input
\usepackage[T1]{fontenc}    % use 8-bit T1 fonts
\usepackage{hyperref}       % hyperlinks
\usepackage{url}            % simple URL typesetting
\usepackage{booktabs}       % professional-quality tables
\usepackage{amsfonts}       % blackboard math symbols
\usepackage{nicefrac}       % compact symbols for 1/2, etc.
\usepackage{microtype}      % microtypography
\usepackage{graphicx}
\usepackage{caption}
\usepackage{mathtools}
\usepackage{subcaption}

\usepackage{multirow}%
\usepackage{amsmath,amssymb}%
\usepackage{amsthm}%
\usepackage{mathrsfs}%
\usepackage[title]{appendix}%
\usepackage{xcolor}%
\usepackage{textcomp}%
\usepackage{manyfoot}%
\usepackage{algorithm}%
\usepackage{algorithmicx}%
\usepackage{algpseudocode}%
\usepackage{listings}%
\usepackage{pgfplots}
\pgfplotsset{compat=1.18}
\usepackage{comment}
\usetikzlibrary{spy}
\usepackage{xfrac}

\usepackage[version=4]{mhchem}
\usepackage{siunitx}
\usepackage{longtable,tabularx}

\usepackage{tikz}
\pgfplotsset{compat=newest}

\definecolor{pv1}{rgb}{0.278431, 0.278431, 0.858824} % Value: 0
\definecolor{pv2}{rgb}{0, 0, 0.360784}               % Value: 0.572
\definecolor{pv3}{rgb}{0, 1, 1}                      % Value: 1.14
\definecolor{pv4}{rgb}{0, 0.501961, 0}               % Value: 1.716
\definecolor{pv5}{rgb}{1, 1, 0}                      % Value: 2.284
\definecolor{pv6}{rgb}{1, 0.380392, 0}               % Value: 2.856
\definecolor{pv7}{rgb}{0.419608, 0, 0}               % Value: 3.428
\definecolor{pv8}{rgb}{0.878431, 0.301961, 0.301961} % Value: 4

\theoremstyle{plain}
\theoremstyle{remark}
\theoremstyle{definition}

\title{Lattice Boltzmann Methods for Compressible (Magneto)hydrodynamics}

\usepackage{amsfonts} 

\author{
  Fedor Bukreev\thanks{These authors contributed equally to this work.} $^,$\thanks{Corresponding author.} \\
  Lattice Boltzmann Research Group (LBRG) \\
  Institute of Mechanical Process Engineering and Mechanics (MVM) \\
  Karlsruhe Institute of Technology (KIT) \\
  Straße am Forum 8, 76131 Karlsruhe, Germany \\
  \texttt{fedor.bukreev@kit.edu} \\
  \And
  Adrian Kummerländer\footnotemark[1] \\
  Lattice Boltzmann Research Group (LBRG) \\
  Institute for Applied and Numerical Mathematics (IANM) \\
  Karlsruhe Institute of Technology (KIT) \\
  Straße am Forum 8 / Englerstraße 2, 76131 Karlsruhe, Germany \\
  \texttt{adrian.kummerlaender@kit.edu} \\
  \And
  Mathias J. Krause\footnotemark[1] \\
  Lattice Boltzmann Research Group (LBRG) \\
  Institute of Mechanical Process Engineering and Mechanics (MVM) \\
  Institute for Applied and Numerical Mathematics (IANM) \\
  Karlsruhe Institute of Technology (KIT) \\
  Straße am Forum 8 / Englerstraße 2, 76131 Karlsruhe, Germany \\
  \texttt{mathias.krause@kit.edu} \\
}

\begin{document}
\maketitle

\begin{abstract}
The simulation of magnetohydrodynamic (MHD) flows presents a highly complex, tightly coupled transport problem that poses severe numerical and computational demands.
Towards this, we propose a novel class of Lattice Boltzmann Methods (LBM) schemes capable of solving a wide range of transport equation systems with high computational efficiency and scalability.
Our approach exploits the algorithmic structure of kinetic formulations to separately transport all state variables of Strang-splitted conservation equations alongside their characteristics, yielding decoupled, fully local operations.

To demonstrate the capability of this framework on complex, numerically demanding multiphysics interactions, we apply it to these MHD flows.
Specifically, we discretize ideal compressible and resistive incompressible MHD systems, which naturally encompass hydrodynamic limits such as the compressible Euler and incompressible Navier-Stokes equations.

Rigorous performance analysis of the implementation within the platform-transparent multi-physics framework OpenLB demonstrates up to 98.9\% of the hardware roofline.
We validate our approach against established incompressible and compressible MHD benchmarks across multiple resolutions.
Finally, we simulate a moving, surface-resolved magnetized asteroid modeled after 16 Psyche in a supersonic early solar wind flow.
This showcases the framework's advanced support for dynamic solid geometries, shifting magnetic fields, and fluid-structure interaction.
\end{abstract}

\keywords{Magnetohydrodynamics \and Lattice Boltzmann Methods \and Compressible \and High-Performance Computing \and Computational Astrophysics}

\section{Introduction}
\emph{Magnetohydrodynamics} (MHD) describes the macroscopic dynamics of electrically conducting fluids, coupling the Navier-Stokes and Maxwell equations~\cite{Liu2017}.
Applications range from liquid metals in industrial applications to highly energetic plasmas in astrophysics.
Simulating these flows poses a severe computational challenge due to strongly non-linear couplings and the strict divergence-free constraint on the magnetic field. Traditional macroscopic solvers rely on complex characteristic decompositions and non-local divergence-cleaning techniques or staggered grids~\cite{Londrillo2000, Gaburov2011}.
As demand for high-fidelity simulations grows, the heavy communication overhead of these non-local operations bottlenecks modern \emph{High-Performance Computing} (HPC) architectures.

The \emph{Lattice Boltzmann Method} (LBM)~\cite{Krueger2016} offers an alternative, decomposing the kinetic algorithm into a perfectly parallel \emph{collision step} and a neighborhood-local \emph{streaming step}.
While ideal for massively parallel execution, extending standard scalar LBM to complex, high-Mach-number multiphysics introduces severe numerical challenges.
Recovering correct macroscopic fluxes often requires complex Hermite expansions or excessive artificial dissipation that compromises physical accuracy~\cite{Wissocq2024}.

To circumvent these limitations, \emph{vectorial LBM} (VLBM) assigns a distinct set of distribution functions to each variable in the macroscopic state vector~\cite{Dubois2014, Wissocq2024}.
Building on this and inspired by unified kinetic schemes~\cite{Liu2017, Grad1949, Struchtrup2003}, our approach treats the fully evolved equation system directly at the mesoscopic level.
Instead of relying on non-local finite differences to approximate spatial derivatives, we apply Strang splitting in time~\cite{Strang1968} and explicitly transport the state variables of the conservation equations alongside their characteristics.
This results in a unified kinetic scheme capable of natively solving a wide range of transport equations, including fully compressible Euler, incompressible Navier-Stokes, and even linear Navier-Cauchy~\cite{Boolakee2025}.

In this work, we present this novel class of LBM schemes and demonstrate its capability on the exceptionally demanding multiphysics interactions of MHD.
By utilizing our fully local formulation, we recover the full compressible macroscopic fluxes while natively handling the divergence-free magnetic field constraint implicitly through localized kinetic updates, bypassing non-local Poisson solvers.

To evaluate the accuracy, robustness, and shock-capturing capabilities of the proposed method, we rigorously validate the scheme against established compressible ideal and incompressible dissipative MHD benchmarks: the Brio-Wu shock tube to demonstrate the method's ability to accurately resolve compound waves, slow-fast shocks, and non-equilibrium kinetic effects at sharp interfaces~\cite{Huang2024}; the MHD rotor problem to assess the handling of strong torsional Alfv\'en waves and rapid rotational discontinuities~\cite{Londrillo2000, Gaburov2011}; and the Orszag-Tang vortex to analyze fidelity in capturing magnetic reconnection, current sheet formation, and the transition to MHD turbulence~\cite{Stone2008, Dellar2013}.
Finally, we highlight the framework's capability for fluid-structure interaction by simulating a complex 3D astrophysical scenario of the solar wind flow interacting with a magnetized asteroid.

The resulting numerical algorithm is implemented and validated in the OpenLB software library~\cite{Krause2021a}, which provides a hardware-abstracted framework capable of highly efficient execution on heterogeneous exascale supercomputers~\cite{Kummerlaender2023, Kummerlaender2026c, Kummerlaender2025}.

In the following, Section~\ref{sec:methodology} introduces a partial differential equation (PDE) agnostic framework for general conservation laws and details our fully local mesoscopic discretization strategy. This methodology is subsequently applied to the governing equations of ideal and resistive magnetohydrodynamics, concluding with the treatment of static and moving boundaries.
In Section~\ref{sec:validation}, we rigorously validate the predictive accuracy and shock-capturing fidelity of our framework against canonical benchmarks.
Section~\ref{sec:performance} evaluates the computational performance and hardware efficiency of our implementation on a modern GPU architecture.
Section~\ref{sec:asteroid} demonstrates the coupled multiphysics capabilities of the framework on a complex 3D computational astrophysics showcase.

\section{Methodology}\label{sec:methodology}

In this section, we present a novel, partial differential equation (PDE) agnostic Lattice Boltzmann framework designed for highly efficient and scalable transport simulations.
Subsequently, we detail the application of this methodology to the target magnetohydrodynamic (MHD) equation systems, including the required evolution of spatial gradients and the realization of boundary conditions for the presented simulation cases.

\subsection{General Conservation Laws}

The proposed numerical framework is designed to solve generic systems of conservation laws that can be written in the abstract vector form:
\begin{align} \label{eq:basic_eq}
    \partial_t \mathbf{Q} + \nabla \cdot \mathbf{\Phi} = \mathbf{S},
\end{align}
where $\mathbf{Q}$ represents an arbitrary macroscopic state vector, $\mathbf{\Phi}$ is the corresponding flux tensor governing the transport of the state variables, and $\mathbf{S}$ encapsulates any localized source or sink terms.
By formulating the numerical method around this generic abstraction, the resulting solver remains largely PDE-agnostic and can be readily adapted to various complex multiphysics problems by redefining the components of $\mathbf{Q}$, $\mathbf{\Phi}$, and $\mathbf{S}$.

\subsection{Lattice Boltzmann Methods}

To discretize the generic conservation laws defined above, we employ a fully local VLBM. 
Unlike finite difference or finite volume methods that rely on extended spatial stencils for flux and gradient evaluation, our approach confines these non-linear operations to strict point-wise updates. 
This algorithmic locality inherently avoids the wide, irregular memory access patterns of traditional solvers, rendering the scheme suited to the saturation of modern heterogeneous HPC architectures.

The fully local VLBM is formulated as:
\begin{align}
    f_{k,i}(\mathbf{x} + \mathbf{c}_i \Delta t, t + \Delta t) = f_{k,i}(\mathbf{x}, t) + \Omega^C \Bigl(\tau_{\text{LB},k}, f_{k,i}(\mathbf{x}, t), f_{k,i}^{eq}(\mathbf{x}, t) \Bigr) + \Omega^S \Bigl(\tau_{\text{LB},k}, f_{k,i}(\mathbf{x}, t), f_{k,i}^{eq}(\mathbf{x}, t),S_{k,i} \Bigr),
\end{align}
where $f_{k,i}$ is the distribution function for a state vector component $k$ in the lattice direction $i$, $\Omega^C$ is the collision operator, $\tau_{\text{LB},k}$ is the lattice relaxation time for a component, $f_{k,i}^{eq}$ is the equilibrium distribution function, and $\mathbf{c}_i = \sfrac{\Delta x}{\Delta t}$ is the lattice velocity vector. $\Omega^S$ is the source operator with the $S_{k,i}$ source term.
The universality of the applied scheme results in the ability to use any possible collision, such as Bhatnagar-Gross-Krook (BGK)~\cite{Bhatnagar1954}, regularized BGK (RLB)~\cite{Latt2006}, two-relaxation-time collision (TRT)~\cite{Ginzburg2018}, multiple relaxation time collision (MRT)~\cite{Coveney2002}, and others, as well as distinct source operators such as direct injection, the Guo scheme~\cite{Guo2002}, the Chai scheme~\cite{Chai2013}, and others. 

In contrast to standard LBM, the non-linear fluxes are embedded directly in the first moment. For this reason, the equilibrium function requires only a linear, first-order expansion:
\begin{align}
    f_{k,i}^{eq} = w_i \left( Q_k + \frac{\mathbf{c}_i \cdot \mathbf{\Phi}_k}{c_s^2} \right),
\end{align}
where $w_i$ are the lattice Gauss-Hermite quadrature weights and $c_s$ is the lattice speed of sound.

To recover the state vector variables and their fluxes, moments of the distribution functions must be taken. The zeroth moment directly recovers the conserved variables, while the first moment exactly recovers the macroscopic fluxes $\mathbf{\Phi}_k$:
\begin{align}
    Q_k = \sum_i f_{k,i}, \quad \quad \mathbf{\Phi}_k = \sum_i f_{k,i} \mathbf{c}_i.
\end{align}

Unlike standard scalar LBM, where the macroscopic fluxes must be recovered from the second-order $\mathcal{O}(\epsilon^2)$ moment expansion, requiring high-order lattice isotropy and complex Hermite polynomials, the present approach strictly embeds the physical non-linear fluxes into the first moment.
This unconditionally guaranties the correct transport equations regardless of the underlying discrete velocity set~\cite{Anandan2024}.
%which is why highly compact lattices like D2Q5 and D3Q7 can be utilized for two-dimensional and three-dimensional simulations (Figure~\ref{fig:lattice} and Table~\ref{tab:latt_par}). 
Furthermore, this PDE-agnostic formulation can be naturally extended to address Navier-Cauchy and other complex multiphysics systems.

For more details and mathematical analysis of VLBM, we recommend the following literature~\cite{Wissocq2024,Boolakee2025,Dubois2014,Anandan2024,Guillon2024}. For the aspects of standard LBM and advection-diffusion equation approximation, please look in~\cite{Krueger2016}.

\paragraph{Lattice relaxation time}
In the case of the standard transport equation approximation, the relaxation time is connected to the diffusion constant $D$ as
\begin{align}
    D=\Bigl(\tau_{\text{LB}} - \frac{1}{2} \Bigr) c_s^2 \Delta t.
\end{align}
The molecular diffusion (or viscous effects) is embedded directly into the flux term, which is why the lattice relaxation time remains only a stabilization parameter.
In the ideal case, $\tau_{\text{LB},k}$ can be set to 0.5, which means zero added artificial diffusion.
Due to the challenge of obtaining a stable simulation with this lattice relaxation time, a stabilizing diffusion is imposed in the following simulations.
By mesh refinement, this stabilization diffusion can be scaled to zero.

\paragraph{Unit conversion}
In contrast to standard LBM, where the SI units are scaled to lattice units with respect to the lattice speed of sound, lattice relaxation time, and a common definition that $\Delta x_{\text{LB}} = \Delta t_{\text{LB}} = 1$, the present approach is no longer bound to this convention.
Identical to other numerical discretization schemes, the unit scaling is performed here with the chosen reference quantities, such as reference length, velocity, viscosity, and others.
This choice is case specific and has only numerical or computational precision limitations.

\subsubsection{Local Moments Evolution}
A fundamental feature of this generic framework is its ability to recover higher moments (stress tensor components, heat fluxes) natively through a local relaxation mechanism, entirely avoiding finite-difference stencils.
When spatial gradients of a variable are required for dissipative fluxes or source terms, they are evolved as independent variables governed by a conservation equation with a relaxation source term. 

Taking a generic spatial gradient tensor $\mathbf{G}$ tracking the gradients of an arbitrary field $\mathbf{V}$ as an example:
\begin{align}
    \partial_t \mathbf{G} + \nabla \cdot \mathbf{\Phi}_{\mathbf{G}} = -\frac{1}{\tau_R} \mathbf{G}.
\end{align}
If analyzed in a quasi-steady state, the time derivative $\partial_t \mathbf{G}$ is negligible. The gradients are thus purely determined by the balance between their flux divergences and their relaxation:
\begin{align}
    \mathbf{G} \approx -\tau_R (\nabla \cdot \mathbf{\Phi}_{\mathbf{G}}).
\end{align}
By formulating the macroscopic flux $\mathbf{\Phi}_{\mathbf{G}}$ such that the field $\mathbf{V}$ is embedded appropriately (e.g., via a $-\mathbf{I} \otimes \mathbf{V}$ term), the spatial divergence operator natively acts upon the field.
Consequently, the relaxation balance recovers the explicitly scaled spatial gradients ($\mathbf{G} \approx \tau_R \nabla \mathbf{V}$) without ever querying a neighboring cell.
Strang splitting~\cite{Strang1968} is applied to handle the calculation of these fluxes from the previous time step.

\subsection{Application to Magnetohydrodynamics}

To demonstrate the capability of this framework on numerically demanding multiphysics interactions, we apply it to compressible and dissipative MHD flows.
The governing equations are formulated by coupling the conservation of mass, momentum, and energy with the magnetic induction equation. They map directly onto the generic target vector form~\ref{eq:basic_eq} whith:
\begin{align} \nonumber
    \mathbf{Q} = \begin{pmatrix} 
    \rho \\ 
    \rho \mathbf{u} \\
    e_{\text{tot}} \\  
    \mathbf{B} 
    \end{pmatrix}, \quad
    \mathbf{\Phi} = \begin{pmatrix*}[l]
    \rho \mathbf{u} \\ 
    \rho \mathbf{u} \otimes \mathbf{u} + p_{\text{tot}}\mathbf{I} - \mathbf{B} \otimes \mathbf{B} - \mathbf{\tau} \\ 
    (e_{\text{tot}} + p_{\text{tot}})\mathbf{u} - \mathbf{B}(\mathbf{u} \cdot \mathbf{B}) - \mathbf{u} \cdot \mathbf{\tau} + \mathbf{q} + \eta \mathbf{j} \times \mathbf{B} \\ 
    \mathbf{u} \otimes \mathbf{B} - \mathbf{B} \otimes \mathbf{u} - \eta \nabla \mathbf{B}
    \end{pmatrix*}, \quad
    \mathbf{S} = \begin{pmatrix} 
    0 \\ 
    \mathbf{0} \\
    0 \\  
    \mathbf{0} 
    \end{pmatrix}. \label{eq::fullMHD}
\end{align}
Here, $\rho$ is density, $\mathbf{u}$ is velocity, $e_{\text{tot}}$ is total energy, $\mathbf{B}$ is the magnetic field, and $p_{\text{tot}} = p + \frac{1}{2}|\mathbf{B}|^2$ is the total pressure. $\mathbf{\tau}$ represents the viscous shear stress tensor, $\mathbf{q}$ is the heat flux, $\eta$ is magnetic diffusivity, $\mathbf{j}$ is electrical density, and $\mathbf{I}$ is the unit tensor. 

Pressure is obtained from the ideal equation of state for compressible gas within a magnetic field:
\begin{align}
    p = (\gamma - 1) \Bigl( e_{\text{tot}} - \frac{1}{2} \rho |\mathbf{u}|^2 -  \frac{1}{2}|\mathbf{B}|^2 \Bigr).
\end{align}

\paragraph{Evolution of the MHD equation system}
In line with the methodology established in the local gradient evolution framework, the macroscopic state vector $\mathbf{Q}$ for the dissipative MHD system is expanded to transport the physical viscous stresses ($\mathbf{\tau}$) and the magnetic field gradients ($\mathbf{G}$) as independent variables:
\begin{align}
    \mathbf{Q} = \begin{pmatrix} 
    \rho \\ 
    \rho \mathbf{u} \\
    e_{\text{tot}} \\  
    \mathbf{B} \\
    \mathbf{\tau} \\
    \mathbf{G}
    \end{pmatrix}.
\end{align}
Here, $\mathbf{G}$ tracks the spatial gradients $\nabla \mathbf{B}$. To recover the physical magnetic field gradients, the macroscopic flux $\mathbf{\Phi}_{\mathbf{G}}$ is formulated as:
\begin{equation}
    \mathbf{\Phi}_{\mathbf{G}} = \mathbf{G} \otimes \mathbf{u} - \mathbf{I} \otimes \mathbf{B}.
\end{equation}
By injecting the magnetic field directly into the flux tensor via the $-\mathbf{I} \otimes \mathbf{B}$ term, the spatial divergence operator natively acts upon the magnetic field ($\nabla \cdot (\mathbf{I} \otimes \mathbf{B}) = \nabla \mathbf{B}$). The relaxation balance under the assumption of small relaxation time $\tau_M$, defined as 0.01 for further simulations, yields $\mathbf{G} \approx \tau_M \nabla \mathbf{B}$. A similar expansion logic is applied to the viscous stress tensor $\mathbf{\tau}$ in the simulations where resistive MHD is considered.

\paragraph{Ideal MHD}
For the most part, we assume the ideal MHD limit where all resistive processes are set to zero, such that viscosity, magnetic diffusivity, and thermal conductivity are null. This leads to a simplified equation system based on the Euler equations:
\begin{align}
    \mathbf{Q} = \begin{pmatrix} 
    \rho \\ 
    \rho \mathbf{u} \\
    e_{\text{tot}} \\  
    \mathbf{B} 
    \end{pmatrix}, \quad
    \mathbf{\Phi} = \begin{pmatrix*}[l] 
    \rho \mathbf{u} \\ 
    \rho \mathbf{u} \otimes \mathbf{u} + p_{\text{tot}}\mathbf{I} - \mathbf{B} \otimes \mathbf{B} \\ 
    (e_{\text{tot}} + p_{\text{tot}})\mathbf{u} - \mathbf{B}(\mathbf{u} \cdot \mathbf{B}) \\ 
    \mathbf{u} \otimes \mathbf{B} - \mathbf{B} \otimes \mathbf{u}
    \end{pmatrix*}.
\end{align}

\paragraph{Incompressible resistive MHD}
For the Dellar's subsonic low compressible Orszag-Tang vortex test~\cite{Dellar2013} (Section~\ref{sec:dellar}), the full equation system is used (Equation~\ref{eq::fullMHD}) with the viscous stress tensor $\mathbf{\tau} = \mu \Delta \mathbf{u}$, where $\mu$ is shear viscosity. Heat fluxes are set to zero, and the magnetic diffusivity is treated as a locally changing function of the current density.

\paragraph{Powell correction}
For the stability of the calculation and the enhancement of the divergence-free magnetic field constraint ($\nabla \cdot \mathbf{B} = 0$), the Powell correction~\cite{Powell1999} is applied as the source terms vector across all shown simulations:
\begin{align}
    \mathbf{S} = -(\nabla \cdot \mathbf{B})
    \begin{pmatrix} 
    0 \\ 
    \mathbf{B} \\
    \mathbf{u} \cdot \mathbf{B} \\  
    \mathbf{u} 
    \end{pmatrix}.
\end{align}

\subsection{Boundary conditions}

In the current work, we use two types of boundary condition conventions: for static boundaries and for the homogenized ones for moving solid objects.

\paragraph{Static boundaries}
All boundary conditions used in further simulations are based on the equilibrium distribution functions. For each state vector variable, the boundaries are either Dirichlet or Neumann. In the first case, the variable value is just defined and in the other case, the needed component is taken from the next cell along the boundary discrete normal.
Having information from the actual and neighboring cells, a new equilibrium distribution function is computed and assigned to the lattice direction.

\paragraph{Homogenization}
For moving solid objects and complex boundaries, we utilize the homogenized LBM (HLBM)~\cite{Krause2017, Kummerlaender2026a, Kummerlaender2026b}.
In standard continuous formulations, this is equivalent to a Brinkman penalization approach~\cite{Angot1999} where a drag term enforces no-slip conditions by driving the permeability to zero inside the solid domain.
In our discrete VLBM framework, instead of applying an explicit Brinkman forcing term, we directly impose the zero permeability limit by blending the state variables.

Specifically, we compute a homogenized target state using a convex combination of the locally computed fluid state and the prescribed solid state.
For MHD, this applies not only to the macroscopic velocity $\mathbf{u}$, but is naturally extended to the magnetic field $\mathbf{B}$:
\begin{align}
    \hat{\mathbf{u}}(\mathbf{x},t) &= d(\mathbf{x},t)\mathbf{u}(\mathbf{x},t) + (1-d(\mathbf{x},t))\mathbf{u}^B(\mathbf{x},t), \\
    \hat{\mathbf{B}}(\mathbf{x},t) &= d(\mathbf{x},t)\mathbf{B}(\mathbf{x},t) + (1-d(\mathbf{x},t))\mathbf{B}^B(\mathbf{x},t),
\end{align}
where $\mathbf{u}^B$ and $\mathbf{B}^B$ are the velocity and magnetic field of the solid object, and the homogenized variables $\hat{\mathbf{u}}$ and $\hat{\mathbf{B}}$ replace the standard variables in the equilibrium distribution and macroscopic flux evaluations.

This homogenization is governed by the lattice porosity parameter $d(\mathbf{x},t) \in [0, 1]$, which depends on the signed distance $\phi(\mathbf{x},t)$ to the solid surface.
Let $\epsilon_h = \epsilon \Delta x$ be the width of the smooth transition region coupled to the grid resolution $\Delta x$.
The porosity is defined as:
\begin{equation}
  d(\mathbf{x},t) := \begin{cases}
    0 & \text{if } \phi(\mathbf{x},t) \le -\frac{\epsilon_h}{2} \\
    s(\phi(\mathbf{x},t)) & \text{if } \phi(\mathbf{x},t) \in \left(-\frac{\epsilon_h}{2}, \frac{\epsilon_h}{2}\right) \\
    1 & \text{if } \phi(\mathbf{x},t) \ge \frac{\epsilon_h}{2}
\end{cases}
\end{equation}
A common choice for the transition function $s$ is the linear profile:
\begin{equation}
  s(\phi) = \frac{\phi}{\epsilon_h} + \frac{1}{2},
\end{equation}
such that the exact solid wall is represented by the level set $d(\mathbf{x},t) = \frac{1}{2}$.

By mapping the macroscopic penalization limits directly to the discrete lattice porosities, fully fluid limit corresponding to $d=1$ and fully solid limit to $d=0$, we bypass the use of an empirical physical permeability model.
While this sharp, single-cell transition introduces an $\mathcal{O}(\epsilon_h)$ truncation error near the interface, the geometric representation strictly converges to a sharp boundary in the limit $\Delta x \to 0$ due to the direct coupling of the transition width to the spatial resolution.

Finally, for zero-gradient boundary conditions on homogenized boundaries, no special algorithmic measures are required; the state vectors, flux vectors, and LBM distribution functions are computed uniformly across the entire domain.

\section{Validation}\label{sec:validation}
In this Section, the validation of the MHD-VLBM is presented on different established benchmarks, showing the robustness and sufficient accuracy of the built solver. In all tests, a D2Q5 lattice velocities set was taken. As the collision operator, we use BGK and add source terms using the direct injection scheme.

\subsection{Brio-Wu shock tube}
The Brio-Wu shock tube is a standard 1D problem performed here in a 2D domain to evaluate the scheme's shock-capturing capabilities in compressible ideal MHD \cite{Huang2024}. The domain is initialized with two distinct states separated by a discontinuity. The left state is given by $(\rho, p, u, v, B_x, B_y) = (1.0, 1.0, 0, 0, 0.75, 1.0)$, and the right state is $(0.125, 0.1, 0, 0, 0.75, -1.0)$ \cite{Brio1988}. The artificial diffusivity was set constantly to $0.75 \cdot 10^{-4}$ $\text{m}^2/\text{s}$. The simulation evolves to produce a complex wave structure consisting of a fast rarefaction wave, a slow compound wave, a contact discontinuity, a slow shock, and a fast rarefaction wave \cite{Huang2024}. This test validates the robustness of the 13-component VLBM in handling strong gradients and assessing non-equilibrium kinetic moments at shock interfaces \cite{Liu2017}.

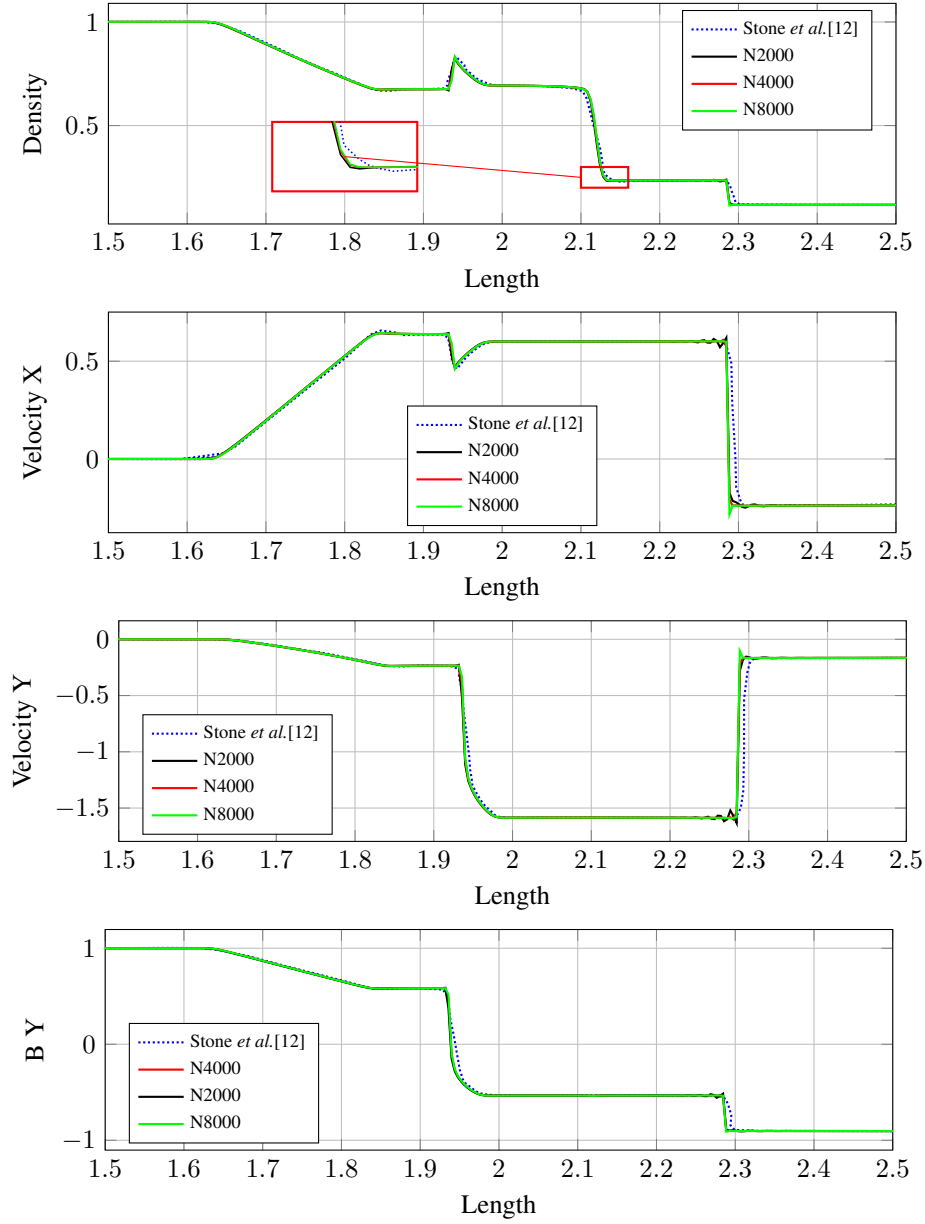
\begin{figure}[htbp]
    \centering
    
    % ==========================================
    % 1. Density Comparison
    % ==========================================
    \begin{tikzpicture}
    \begin{axis}[
        name=mainplot,
        width=12cm,
        height=4.5cm,
        xlabel={Length},
        ylabel={Density},
        legend pos=north east,
        grid=major,
        legend cell align={left},
        xmin=1.5, 
        xmax=2.5,
        legend style={font=\scriptsize}
    ]
        % Reference Data
        \addplot[mark=none, thick, densely dotted, blue] table [x expr=\thisrow{x} + 2, y index=1] {brio_wu_shock_tube/dens_ref.csv};
        \addlegendentry{Stone~\textit{et al.}\cite{Stone2008}}
        
        % Simulated Data
        \addplot[mark=none, thick, black] table [x=Points_0, y=Density] {brio_wu_shock_tube/N2000_D2Q5.csv};
        \addlegendentry{N2000}

        \addplot[mark=none, thick, red] table [x=Points_0, y=Density] {brio_wu_shock_tube/N4000_D2Q5.csv};
        \addlegendentry{N4000}

        \addplot[mark=none, thick, green] table [x=Points_0, y=Density] {brio_wu_shock_tube/N8000_D2Q5.csv};
        \addlegendentry{N8000}

        % --- Target Box Formatting (Red - Around (2.13, 0.25)) ---
        % You can adjust these bounds if you want to capture more/less of the wave
        \draw[red, thick] (axis cs:2.10, 0.20) rectangle (axis cs:2.16, 0.30);
        
        % Inset placement and connecting point
        \coordinate (insetPosition) at (axis cs:1.8, 0.35); 
        \coordinate (boxEdge) at (axis cs:2.10, 0.25);
    \end{axis}

    % ==========================================
    % INSET PLOT (Zoomed Region)
    % ==========================================
    \begin{axis}[
        at={(insetPosition)},
        anchor=center,
        width=3.5cm,  
        height=2.5cm, 
        xmin=2.10,   
        xmax=2.16,
        ymin=0.20,    
        ymax=0.30,
        axis background/.style={fill=white}, 
        axis line style={red, thick},        
        xtick=\empty, 
        ytick=\empty,
        enlargelimits=false
    ]
        % Re-plot all curves for the zoomed region (semithick for clarity)
        \addplot[mark=none, semithick, densely dotted, blue] table [x expr=\thisrow{x} + 2, y index=1] {brio_wu_shock_tube/dens_ref.csv};
        \addplot[mark=none, semithick, black] table [x=Points_0, y=Density] {brio_wu_shock_tube/N2000_D2Q5.csv};
        \addplot[mark=none, semithick, red] table [x=Points_0, y=Density] {brio_wu_shock_tube/N4000_D2Q5.csv};
        \addplot[mark=none, semithick, green] table [x=Points_0, y=Density] {brio_wu_shock_tube/N8000_D2Q5.csv};
    \end{axis}

    % Connecting Line
    \draw[red] (boxEdge) -- (insetPosition);
\end{tikzpicture}

    \vspace{0.1cm}

    % ==========================================
    % 2. Velocity Comparison
    % ==========================================
    \begin{tikzpicture}
        \begin{axis}[
            width=12cm,
            height=4.5cm,
            xlabel={Length},
            ylabel={Velocity X},
            legend style={at={(0.5,0.03)}, anchor=south, font=\scriptsize},
            grid=major,
            legend cell align={left},
            xmin=1.5, 
            xmax=2.5
        ]
            % Reference Data
            \addplot[mark=none, thick, densely dotted, blue] table [x expr=\thisrow{x} + 2, y index=1] {brio_wu_shock_tube/u_x_ref.csv};
            \addlegendentry{Stone~\textit{et al.}\cite{Stone2008}}
            
            % Simulated Data
            \addplot[mark=none, thick, black] table [x=Points_0, y=Velocity_0] {brio_wu_shock_tube/N2000_D2Q5.csv};
            \addlegendentry{N2000}

            \addplot[mark=none, thick, red] table [x=Points_0, y=Velocity_0] {brio_wu_shock_tube/N4000_D2Q5.csv};
            \addlegendentry{N4000}

            \addplot[mark=none, thick, green] table [x=Points_0, y=Velocity_0] {brio_wu_shock_tube/N8000_D2Q5.csv};
            \addlegendentry{N8000}
        \end{axis}
    \end{tikzpicture}

    \vspace{0.1cm}

    % ==========================================
    % 3. Velocity Y Comparison
    % ==========================================
    \begin{tikzpicture}
        \begin{axis}[
            width=12cm,
            height=4.5cm,
            xlabel={Length},
            ylabel={Velocity Y},
            legend pos=south west,
            grid=major,
            legend cell align={left},
            xmin=1.5, 
            xmax=2.5,
            legend style={font=\scriptsize}
        ]
            % Reference Data
            \addplot[mark=none, thick, densely dotted, blue] table [x expr=\thisrow{x} + 2, y index=1] {brio_wu_shock_tube/u_y_ref.csv};
            \addlegendentry{Stone~\textit{et al.}\cite{Stone2008}}
            
            % Simulated Data
            \addplot[mark=none, thick, black] table [x=Points_0, y=Velocity_1] {brio_wu_shock_tube/N2000_D2Q5.csv};
            \addlegendentry{N2000}

            \addplot[mark=none, thick, red] table [x=Points_0, y=Velocity_1] {brio_wu_shock_tube/N4000_D2Q5.csv};
            \addlegendentry{N4000}

            \addplot[mark=none, thick, green] table [x=Points_0, y=Velocity_1] {brio_wu_shock_tube/N8000_D2Q5.csv};
            \addlegendentry{N8000}
        \end{axis}
    \end{tikzpicture}

    \vspace{0.1cm}

    % ==========================================
    % 4. Magnetic Field Comparison
    % ==========================================
    \begin{tikzpicture}
        \begin{axis}[
            width=12cm,
            height=4.5cm,
            xlabel={Length},
            ylabel={B Y},
            legend pos=south west,
            grid=major,
            legend cell align={left},
            xmin=1.5, 
            xmax=2.5,
            legend style={font=\scriptsize}
        ]
            % Reference Data
            \addplot[mark=none, thick, densely dotted, blue] table [x expr=\thisrow{x} + 2, y index=1] {brio_wu_shock_tube/B_y_ref.csv};
            \addlegendentry{Stone~\textit{et al.}\cite{Stone2008}}
            
            % Simulated Data
            \addplot[mark=none, thick, red] table [x=Points_0, y=MagneticField_1] {brio_wu_shock_tube/N4000_D2Q5.csv};
            \addlegendentry{N4000}

            \addplot[mark=none, thick, black] table [x=Points_0, y=MagneticField_1] {brio_wu_shock_tube/N2000_D2Q5.csv};
            \addlegendentry{N2000}

            \addplot[mark=none, thick, green] table [x=Points_0, y=MagneticField_1] {brio_wu_shock_tube/N8000_D2Q5.csv};
            \addlegendentry{N8000}
        \end{axis}
    \end{tikzpicture}
    
    \caption{Comparison of reference and simulated data for Brio-Wu benchmark at $t=0.2$.}
    \label{fig:brio_wu}
\end{figure}

It can be seen in Figure~\ref{fig:brio_wu} that the results of different resolutions (2000, 4000, 8000 cells in the domain length), even without separate shock capturing, show good accordance with the reference solution. At finer cell sizes, the oscillations disappear and the variables become smoother. This test shows the validity of the developed solver for accurate reproduction of shock fronts bound to magnetic field in one dimension. 

\subsection{MHD rotor}
The MHD rotor problem tests the propagation of strong torsional Alfv\'en waves in a compressible medium~\cite{Londrillo2000}. The setup consists of a rapidly rotating, dense fluid cylinder embedded in a uniform static background fluid, both permeated by a constant horizontal magnetic field $B_x = 5/\sqrt{4\pi}$ (with $B_y = 0$). Specifically, within a fully periodic $1 \times 1$ spatial domain, a rotor of radius $r_0 = 0.1$ with density $\rho=10.0$ and initial velocity fields $u = -2y/r_0$ and $v = 2x/r_0$ is placed in a resting background medium ($u=v=0$) with $\rho=1.0$. The initial thermodynamic pressure is uniformly set to $p=1.0$ across the domain, and the fluid is modeled as an ideal gas with an adiabatic index of $\gamma=1.4$. To reduce initial extreme transients, the density and velocity fields are smoothly connected between the core and the background via a linear taper from $r_0 = 0.1$ to $r_1 = 0.115$. The strong rotational kinetic energy twists the magnetic field lines, launching outward-propagating Alfv\'en waves that decelerate the rotor while compressing the surrounding fluid~\cite{Gaburov2011}. 

For this specific configuration, the VLBM simulation is performed at resolutions of $500 \times 500$, $1000 \times 1000$, and $2000 \times 2000$. The simulation runs up to a final time of $t = 0.16$, utilizing a CFL number of $0.01$ (resulting in a physical time step of $\Delta t \approx 1.3 \times 10^{-6}$). In the simulation, the artificial diffusivity of $3 \cdot 10^{-4}, 1.5 \cdot 10^{-4}, 0.75 \cdot 10^{-4}$ $\text{m}^2/\text{s}$ is used respectively to ensure stability during the robust flow transients.

\begin{figure}[htbp]
    \centering

    % --- Left Plot: Density ---
    \begin{minipage}{0.48\linewidth}
        \centering
        \def\minval{0}
        \def\maxval{13}

        \begin{tikzpicture}[x=0.8\linewidth]
            % Draw Paraview gradient based on relative positions (x / 4)
            \shade[left color=pv1, right color=pv2] (0,0) rectangle (0.143, 0.2);
            \shade[left color=pv2, right color=pv3] (0.143,0) rectangle (0.285, 0.2);
            \shade[left color=pv3, right color=pv4] (0.285,0) rectangle (0.429, 0.2);
            \shade[left color=pv4, right color=pv5] (0.429,0) rectangle (0.571, 0.2);
            \shade[left color=pv5, right color=pv6] (0.571,0) rectangle (0.714, 0.2);
            \shade[left color=pv6, right color=pv7] (0.714,0) rectangle (0.857, 0.2);
            \shade[left color=pv7, right color=pv8] (0.857,0) rectangle (1, 0.2);
            \draw[black] (0,0) rectangle (1, 0.2); % Outline the whole bar

            % Loop for major ticks based on new scale
            \foreach \val in {0, 3, 6, 9, 13} {
                % Calculate exact position from 0 to 1 based on the physical value
                \pgfmathsetmacro{\xpos}{(\val - \minval) / (\maxval - \minval)}
                \draw (\xpos, 0.2) -- (\xpos, 0.25); % Small tick mark
                \node[above, font=\footnotesize] at (\xpos, 0.25) {\val};
            }
        \end{tikzpicture}

        \vspace{1mm}
        \includegraphics[width=\linewidth]{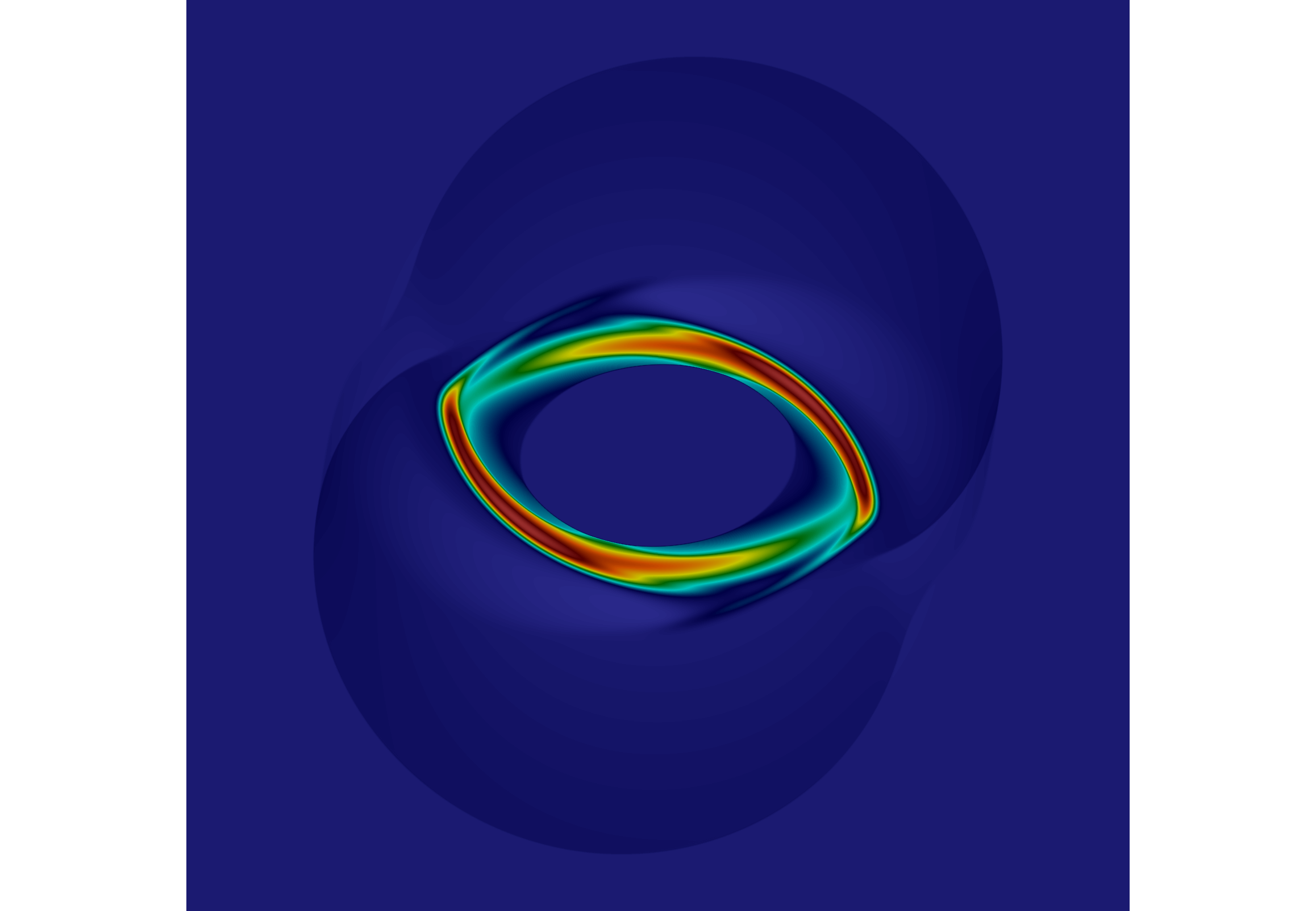}
    \end{minipage}
    \hspace{3mm} % Fixed, small spacing to move panels together
    % --- Right Plot: Mach number ---
    \begin{minipage}{0.48\linewidth}
        \centering
        \def\minval{0}
        \def\maxval{4}

        \begin{tikzpicture}[x=0.8\linewidth]
            % Draw Paraview gradient based on relative positions (x / 4)
            \shade[left color=pv1, right color=pv2] (0,0) rectangle (0.143, 0.2);
            \shade[left color=pv2, right color=pv3] (0.143,0) rectangle (0.285, 0.2);
            \shade[left color=pv3, right color=pv4] (0.285,0) rectangle (0.429, 0.2);
            \shade[left color=pv4, right color=pv5] (0.429,0) rectangle (0.571, 0.2);
            \shade[left color=pv5, right color=pv6] (0.571,0) rectangle (0.714, 0.2);
            \shade[left color=pv6, right color=pv7] (0.714,0) rectangle (0.857, 0.2);
            \shade[left color=pv7, right color=pv8] (0.857,0) rectangle (1, 0.2);
            \draw[black] (0,0) rectangle (1, 0.2); % Outline the whole bar

            % Simplified loop for labels from 0 to 4
            \foreach \val in {0, 1, 2, 3, 4} {
                % Calculate exact position from 0 to 1 based on the physical value
                \pgfmathsetmacro{\xpos}{(\val - \minval) / (\maxval - \minval)}
                \draw (\xpos, 0.2) -- (\xpos, 0.25); % Small tick mark
                \node[above, font=\footnotesize] at (\xpos, 0.25) {\val};
            }
        \end{tikzpicture}

        \vspace{1mm}
        \includegraphics[width=\linewidth]{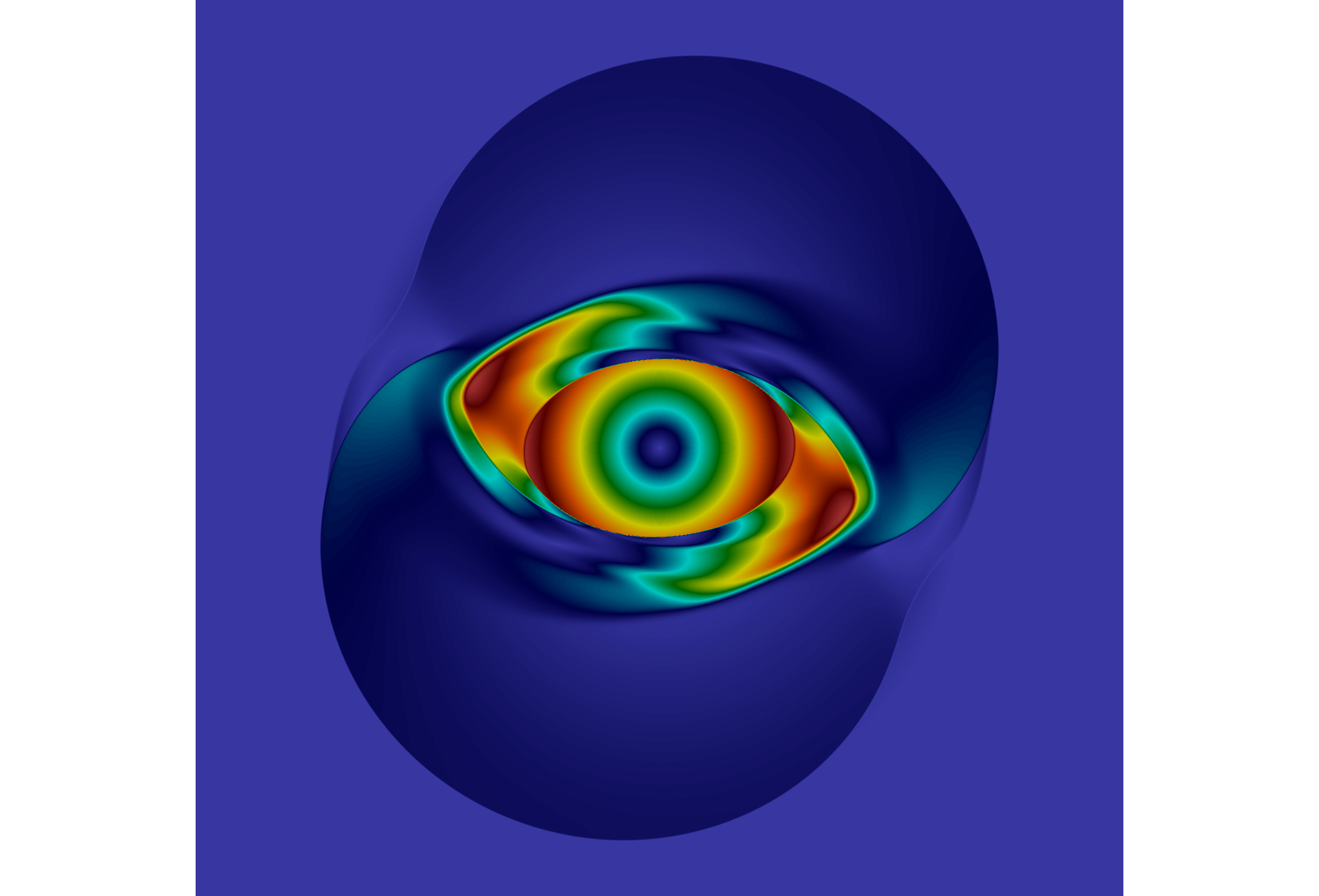}
    \end{minipage}

    % Updated caption
    \caption{Density and local Mach number distribution at $t=0.15$ by resolution of 2000, Results can be compared to Stone~\textit{et al.}~\cite{Stone2008}.}
    \label{fig:rotor}
\end{figure}

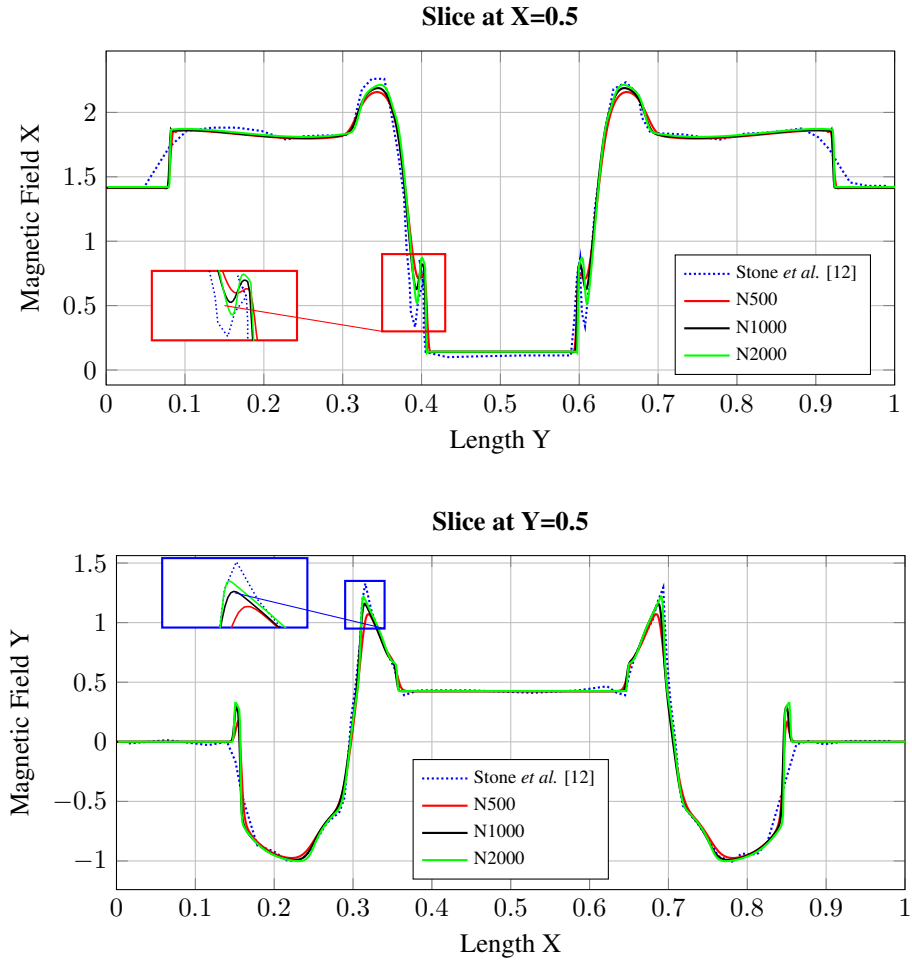
\begin{figure}[htbp]
    \centering

    % ==========================================
    % PLOT 1: SLICE AT X=0.5 (with Red Inset)
    % ==========================================
    \begin{tikzpicture}
        \begin{axis}[
            name=plot1,
            width=12cm,
            height=6cm,
            title={\textbf{Slice at X=0.5}},
            xlabel={Length Y},
            ylabel={Magnetic Field X},
            xmin = 0,
            xmax = 1,
            legend pos=south east,
            grid=major,
            legend cell align={left},
            legend style={font=\scriptsize},
            ytick distance=0.5,
        ]
            % Reference Data
            \addplot[mark=none, thick, densely dotted, blue] table [x index=0, y index=1] {MHD_rotor/B_x_ref.csv};
            \addlegendentry{Stone~\textit{et al.} \cite{Stone2008}}
            
            % Simulated Data
            \addplot[mark=none, thick, red] table [x=arc_length, y=MagneticField_0] {MHD_rotor/N500_D2Q5_CFL0_01_factor0_5_y_line.csv};
            \addlegendentry{N500}

            \addplot[mark=none, thick, black] table [x=arc_length, y=MagneticField_0] {MHD_rotor/N1000_D2Q5_CFL0_01_factor0_25_y_line.csv};
            \addlegendentry{N1000}

            \addplot[mark=none, thick, green] table [x=arc_length, y=MagneticField_0] {MHD_rotor/N2000_D2Q5_CFL0_01_factor0_125_y_line.csv};
            \addlegendentry{N2000}

            % --- Target Box 1 Formatting (Red - Around (0.08, 1.85)) ---
            \draw[red, thick] (axis cs:0.35, 0.3) rectangle (axis cs:0.43, 0.9);
            
            \coordinate (insetPosition1) at (axis cs:0.15, 0.5); 
            \coordinate (boxEdge1) at (axis cs:0.35, 0.3);    
        \end{axis}

        % --- INSET PLOT 1 ---
        \begin{axis}[
            at={(insetPosition1)},
            anchor=center,
            width=3.5cm,  
            height=2.5cm, 
            xmin=0.35,   
            xmax=0.43,
            ymin=0.3,    
            ymax=0.9,
            axis background/.style={fill=white}, 
            axis line style={red, thick},        
            xtick=\empty, 
            ytick=\empty,
            enlargelimits=false
        ]
            \addplot[mark=none, semithick, densely dotted, blue] table [x index=0, y index=1] {MHD_rotor/B_x_ref.csv};
            \addplot[mark=none, semithick, red] table [x=arc_length, y=MagneticField_0] {MHD_rotor/N500_D2Q5_CFL0_01_factor0_5_y_line.csv};
            \addplot[mark=none, semithick, black] table [x=arc_length, y=MagneticField_0] {MHD_rotor/N1000_D2Q5_CFL0_01_factor0_25_y_line.csv};
            \addplot[mark=none, semithick, green] table [x=arc_length, y=MagneticField_0] {MHD_rotor/N2000_D2Q5_CFL0_01_factor0_125_y_line.csv};
        \end{axis}

        % Connecting Line 1
        \draw[red] (boxEdge1) -- (insetPosition1);
    \end{tikzpicture}

    \vspace{0.5cm}

    % ==========================================
    % PLOT 2: SLICE AT Y=0.5 (with Blue Inset)
    % ==========================================
    \begin{tikzpicture}
        \begin{axis}[
            name=plot2,
            width=12cm,
            height=6cm,
            title={\textbf{Slice at Y=0.5}},
            xlabel={Length X},
            ylabel={Magnetic Field Y},
            xmin = 0,
            xmax = 1,
            legend style={at={(0.5,0.03)}, anchor=south, font=\scriptsize},
            grid=major,
            legend cell align={left},
            ytick distance=0.5
        ]
            % Reference Data
            \addplot[mark=none, thick, densely dotted, blue] table [x index=0, y index=1] {MHD_rotor/B_y_ref.csv};
            \addlegendentry{Stone~\textit{et al.} \cite{Stone2008}}
            
            % Simulated Data
            \addplot[mark=none, thick, red] table [x=arc_length, y=MagneticField_1] {MHD_rotor/N500_D2Q5_CFL0_01_factor0_5_x_line.csv};
            \addlegendentry{N500}

            \addplot[mark=none, thick, black] table [x=arc_length, y=MagneticField_1] {MHD_rotor/N1000_D2Q5_CFL0_01_factor0_25_x_line.csv};
            \addlegendentry{N1000}

            \addplot[mark=none, thick, green] table [x=arc_length, y=MagneticField_1] {MHD_rotor/N2000_D2Q5_CFL0_01_factor0_125_x_line.csv};
            \addlegendentry{N2000}

            % --- Target Box 2 Formatting (Blue - Around (0.33, 1.25)) ---
            \draw[blue, thick] (axis cs:0.29, 0.95) rectangle (axis cs:0.34, 1.35);
            
            \coordinate (insetPosition2) at (axis cs:0.15, 1.25); 
            \coordinate (boxEdge2) at (axis cs:0.34, 0.95);
        \end{axis}

        % --- INSET PLOT 2 ---
        \begin{axis}[
            at={(insetPosition2)},
            anchor=center,
            width=3.5cm,  
            height=2.5cm, 
            xmin=0.29,   
            xmax=0.34,
            ymin=0.95,    
            ymax=1.35,
            axis background/.style={fill=white}, 
            axis line style={blue, thick},        
            xtick=\empty, 
            ytick=\empty,
            enlargelimits=false
        ]
            \addplot[mark=none, semithick, densely dotted, blue] table [x index=0, y index=1] {MHD_rotor/B_y_ref.csv};
            \addplot[mark=none, semithick, red] table [x=arc_length, y=MagneticField_1] {MHD_rotor/N500_D2Q5_CFL0_01_factor0_5_x_line.csv};
            \addplot[mark=none, semithick, black] table [x=arc_length, y=MagneticField_1] {MHD_rotor/N1000_D2Q5_CFL0_01_factor0_25_x_line.csv};
            \addplot[mark=none, semithick, green] table [x=arc_length, y=MagneticField_1] {MHD_rotor/N2000_D2Q5_CFL0_01_factor0_125_x_line.csv};
        \end{axis}

        % Connecting Line 2
        \draw[blue] (boxEdge2) -- (insetPosition2);
    \end{tikzpicture}
    
    \caption{Behavior of magnetic field X at the x = 0.5 slice (top panel) and magnetic field Y in y = 0.5 slice (bottom) for the MHD rotor problem at t = 0.15.}
    \label{fig:mhd_rotor}
\end{figure}

In the Figure~\ref{fig:rotor}, density and local Mach number distributions show smooth shock fronts, matching corresponding figures in the reference work~\cite{Stone2008}. The plots in Figure~\ref{fig:mhd_rotor} show good agreement with the spectral solution, which is accentuated in the subplots. They underline the convergent behavior of the refinement with acoustic scaling and reducing stabilization diffusivity. As we don’t apply a separate shock capturer in these validations, at the ends of the domain, the shocks are sharper than the goal reference because of the oscillations occurring there. In future works, we will investigate shock capturing techniques to handle such situations involving problems with discontinuities.

\subsection{Low compressible resistive Orszag-Tang vortex} \label{sec:dellar}
The Orszag-Tang vortex is a canonical benchmark for studying the transition to 2D MHD turbulence, characterized by strong velocity-magnetic field coupling and current-sheet formation. To study the incompressible limit, the initial conditions are prescribed via a fluid streamfunction $\phi = 2\cos(x) - 2\sin(y)$ and a magnetic flux function $\psi = 2\cos(x) - \cos(2y)$, which yield the velocity profiles $\mathbf{u} = [2\cos(y), -2\sin(x)]$ and magnetic profiles $\mathbf{B} = [-2\sin(2y), -2\sin(x)]$. The system is initialized with a uniform density $\rho_0=1$ and a sufficiently small Mach number to suppress spurious compressibility effects~\cite{Dellar2013}. The evolution is tracked using a kinematic viscosity of $\nu = 10^{-3}$ (corresponding to an effective Reynolds number of $Re = 4000\pi$), monitoring the peak electric current density and vorticity to evaluate the behavior of a current-dependent resistivity model $\eta(j)=\eta_0 (1 + (\sfrac{j}{j_c})^2)$ as the system generates complex, fine-scale turbulent structures. $\eta_0$ is set equal to the kinematic viscosity, and $j_c = 25$. To match the results of Dellar, the stabilization diffusivity was set to 5\% of the physical viscosity.

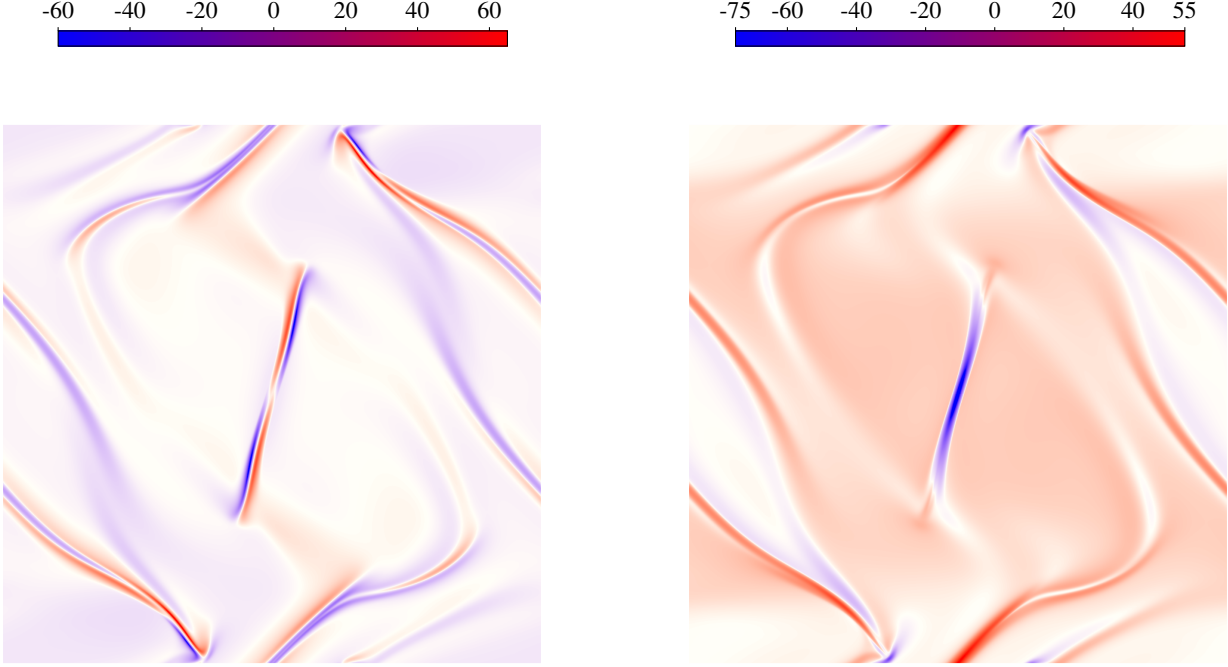
\begin{figure}[htbp]
    \centering
    
    % --- Left Plot: Vorticity ---
    \begin{minipage}{0.45\linewidth}
        \centering
        \def\minval{-60}
        \def\maxval{65}
        
        \begin{tikzpicture}[x=0.8\linewidth] 
            % Draw standard Blue-White-Red gradient (white is forced exactly at 50% width)
            \shade[left color=blue, middle color=white, right color=red, draw=black] (0,0) rectangle (1, 0.2);
            
            % Loop for labels every 20 units
            \foreach \val in {-60, -40, -20, 0, 20, 40, 60} {
                % Calculate exact position from 0 to 1 based on the physical value
                \pgfmathsetmacro{\xpos}{(\val - \minval) / (\maxval - \minval)}
                \draw (\xpos, 0.2) -- (\xpos, 0.25); % Small tick mark
                \node[above, font=\footnotesize] at (\xpos, 0.25) {\val};
            }
        \end{tikzpicture}

        \vspace{1mm}
        \includegraphics[width=\linewidth]{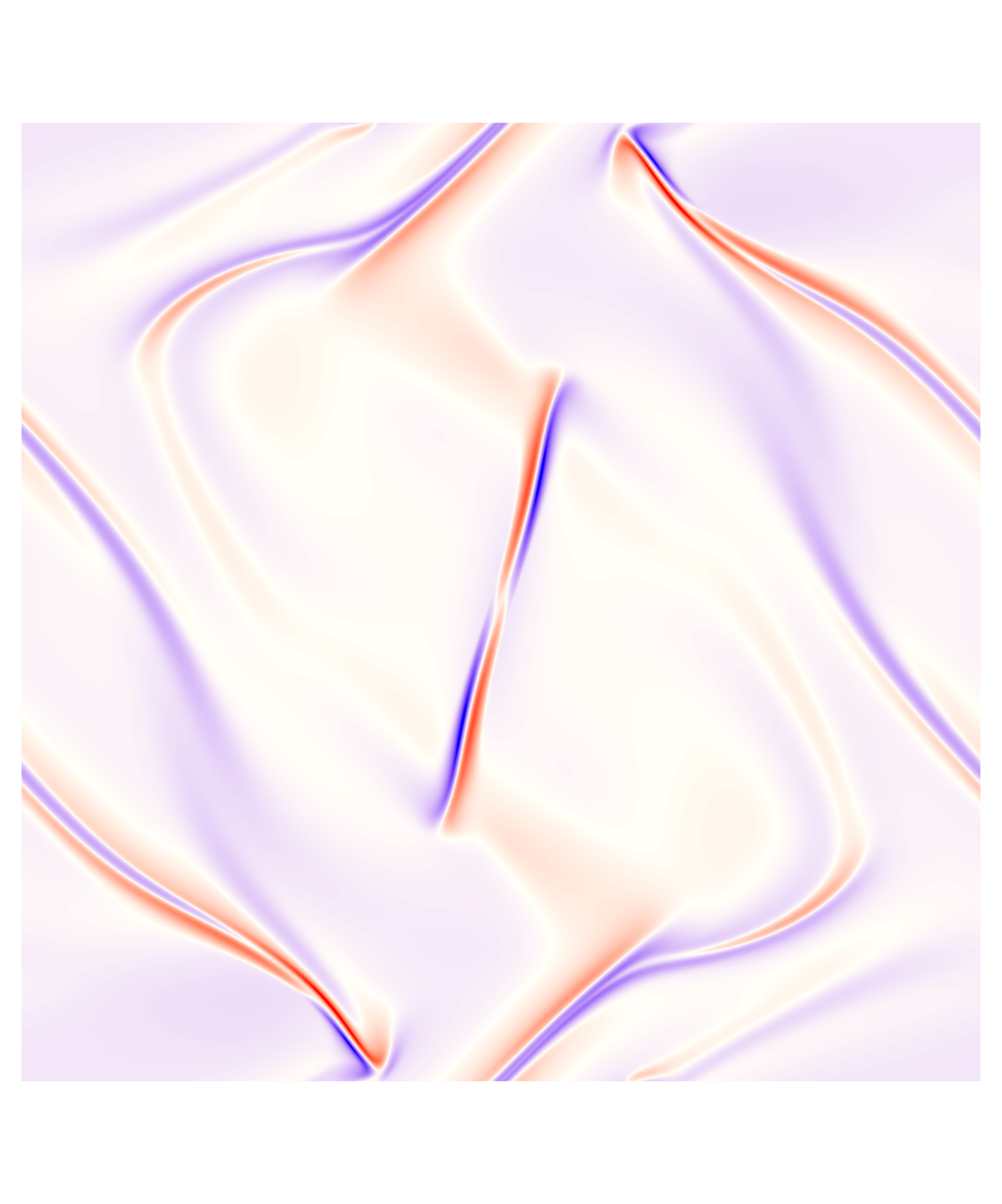}
    \end{minipage}
    \hfill 
    % --- Right Plot: Current Density ---
    \begin{minipage}{0.45\linewidth}
        \centering
        \def\minval{-75}
        \def\maxval{55}
        
        \begin{tikzpicture}[x=0.8\linewidth] 
            % Draw standard Blue-White-Red gradient (white is forced exactly at 50% width)
            \shade[left color=blue, middle color=white, right color=red, draw=black] (0,0) rectangle (1, 0.2);
            
            % Add absolute min value (-75)
            \draw (0, 0.2) -- (0, 0.25);
            \node[above, font=\footnotesize] at (0, 0.25) {\minval};
            
            % Loop for labels every 20 units (aligned to 0)
            \foreach \val in {-60, -40, -20, 0, 20, 40} {
                % Calculate exact position from 0 to 1 based on the physical value
                \pgfmathsetmacro{\xpos}{(\val - \minval) / (\maxval - \minval)}
                \draw (\xpos, 0.2) -- (\xpos, 0.25); % Small tick mark
                \node[above, font=\footnotesize] at (\xpos, 0.25) {\val};
            }
            
            % Add absolute max value (55)
            \draw (1, 0.2) -- (1, 0.25);
            \node[above, font=\footnotesize] at (1, 0.25) {\maxval};
        \end{tikzpicture}

        \vspace{1mm}
        \includegraphics[width=\linewidth]{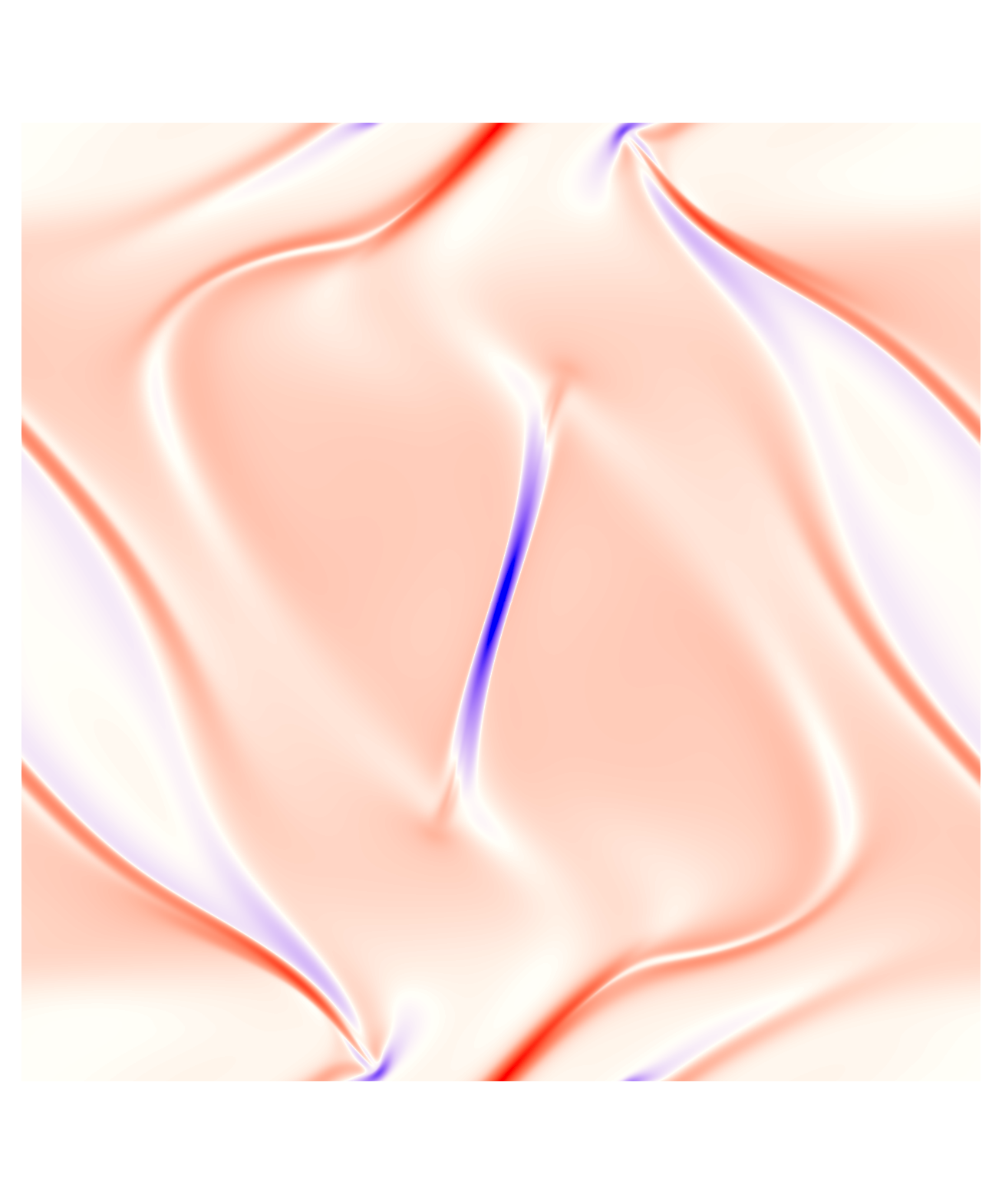}
    \end{minipage}

    \caption{Vorticity and electric current density by resolution of 2048 cells in the side length at $t=1$ by incompressible Orszag-Tang vortex. Results can be compared to Dellar~\cite{Dellar2013}.}
    \label{fig:dellar}
\end{figure}

\begin{figure}[htb]
    \centering
    \begin{tikzpicture}
        % ==========================================
        % 1. MAIN PLOT
        % ==========================================
        \begin{axis}[
            name=mainplot,
            width=12cm,
            height=6cm,
            title={\textbf{Y=$\mathbf{\pi}$}},
            xlabel={Length X},
            ylabel={Electric current density},
            xmin = 0,
            xmax = 1,
            legend pos=south east,
            grid=major,
            legend cell align={left},
            ytick distance=20,
            legend style={font=\scriptsize}
        ]

            % Reference Data
            \addplot[mark=none, thick, densely dotted, blue] table [x index=0, y index=1] {low_Mach_Dellar/j_Dellar.csv};
            \addlegendentry{Dellar~\cite{Dellar2013}}

            % Simulated Data N512
            \addplot[mark=none, thick, red] table [x expr=\thisrow{arc_length} / 6.275, y=ElectricCurrentDensity] {low_Mach_Dellar/N512_D2Q5.csv};
            \addlegendentry{N512}

            % Simulated Data N1024
            \addplot[mark=none, thick, black] table [x expr=\thisrow{arc_length} / 6.275, y=ElectricCurrentDensity] {low_Mach_Dellar/N1024_D2Q5.csv};
            \addlegendentry{N1024}

            % Simulated Data N2048
            \addplot[mark=none, thick, green] table [x expr=\thisrow{arc_length} / 6.2833, y=ElectricCurrentDensity] {low_Mach_Dellar/N2048_D2Q5.csv};
            \addlegendentry{N2048}

            % --- Target Box 1 Formatting (Red - Central Dip) ---
            \draw[red, thick] (axis cs:0.47, -5) rectangle (axis cs:0.5, 20);
            
            \coordinate (insetPosition1) at (axis cs:0.2, -30); 
            \coordinate (boxEdge1) at (axis cs:0.47, -5);    

            % --- Target Box 2 Formatting (Blue - Side Peak) ---
            \draw[blue, thick] (axis cs:0.88, -20) rectangle (axis cs:0.92, 0);
            
            \coordinate (insetPosition2) at (axis cs:0.63, -30); 
            \coordinate (boxEdge2) at (axis cs:0.88, -10);

        \end{axis}

        % ==========================================
        % 2. INSET PLOT 1 (Red - Central Dip)
        % ==========================================
        \begin{axis}[
            at={(insetPosition1)},
            anchor=center,
            width=3.5cm,  
            height=2.5cm, 
            xmin=0.47,   
            xmax=0.5,
            ymin=-10,    
            ymax=20,
            axis background/.style={fill=white}, 
            axis line style={red, thick},        
            xtick=\empty, 
            ytick=\empty,
            enlargelimits=false
        ]
            \addplot[mark=none, semithick, densely dotted, blue] table [x index=0, y index=1] {low_Mach_Dellar/j_Dellar.csv};
            \addplot[mark=none, semithick, red] table [x expr=\thisrow{arc_length} / 6.275, y=ElectricCurrentDensity] {low_Mach_Dellar/N512_D2Q5.csv};
            \addplot[mark=none, semithick, black] table [x expr=\thisrow{arc_length} / 6.275, y=ElectricCurrentDensity] {low_Mach_Dellar/N1024_D2Q5.csv};
            \addplot[mark=none, semithick, green] table [x expr=\thisrow{arc_length} / 6.2833, y=ElectricCurrentDensity] {low_Mach_Dellar/N2048_D2Q5.csv};
        \end{axis}

        % ==========================================
        % 3. INSET PLOT 2 (Blue - Side Peak)
        % ==========================================
        \begin{axis}[
            at={(insetPosition2)},
            anchor=center,
            width=3.5cm,  
            height=3.5cm, 
            xmin=0.88,   
            xmax=0.92,
            ymin=-20,    
            ymax=0,
            axis background/.style={fill=white}, 
            axis line style={blue, thick},        
            xtick=\empty, 
            ytick=\empty,
            enlargelimits=false
        ]
            \addplot[mark=none, semithick, densely dotted, blue] table [x index=0, y index=1] {low_Mach_Dellar/j_Dellar.csv};
            \addplot[mark=none, semithick, red] table [x expr=\thisrow{arc_length} / 6.275, y=ElectricCurrentDensity] {low_Mach_Dellar/N512_D2Q5.csv};
            \addplot[mark=none, semithick, black] table [x expr=\thisrow{arc_length} / 6.275, y=ElectricCurrentDensity] {low_Mach_Dellar/N1024_D2Q5.csv};
            \addplot[mark=none, semithick, green] table [x expr=\thisrow{arc_length} / 6.2833, y=ElectricCurrentDensity] {low_Mach_Dellar/N2048_D2Q5.csv};
        \end{axis}

        % ==========================================
        % 4. CONNECTING LINES
        % ==========================================
        \draw[red] (boxEdge1) -- (insetPosition1);
        \draw[blue] (boxEdge2) -- (insetPosition2);

    \end{tikzpicture}
    \caption{Electric current density $j$ along the horizontal line at the height of $y=\pi$ compared to the spectral simulation taken from~\cite{Dellar2013}.}
    \label{fig:mhd_dellar}
\end{figure}
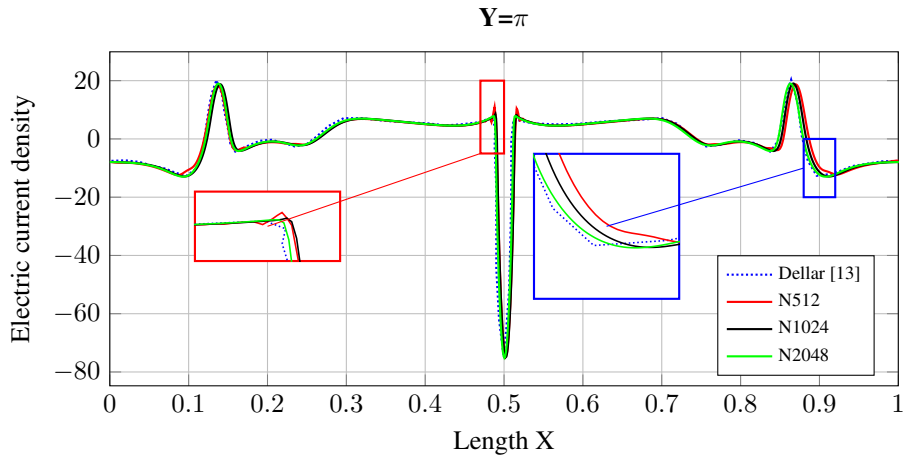

The visual plots (Figure~\ref{fig:dellar}) match those in the referenced publication and show smooth vorticity and electric current density distributions in the domain. The Figure~\ref{fig:mhd_dellar} shows how the values are converging towards the spectral simulation values through acoustic spatiotemporal scaling with constant stabilization diffusion. It can be seen that the numerical diffusion is reducing with the refinement of the cell size, and the finest resolution matches the reference spectral values as well as the standard LBM simulation results of~\cite{Dellar2013}.

\subsection{Compressible non-resistive Orszag-Tang vortex}
To evaluate the scheme's ability to capture complex compressible dynamics and multiple shock formations, the adiabatic non-resistive Orszag-Tang vortex for ideal MHD is simulated following the standard validation setup of \cite{Stone2008} and \cite{Gaburov2011}. The problem is initialized on a periodic domain of $[0, 1] \times [0, 1]$ with an adiabatic equation of state ($\gamma = 5/3$). The initial density and pressure are set uniformly to $\rho = \frac{25}{36\pi}$ and $p = \frac{5}{12\pi}$, yielding a nominal sound speed of $c_s = 1.0$. 

The velocity and magnetic field profiles are defined as $\mathbf{u} = [ -\sin(2\pi y), \sin(2\pi x)]$, $\mathbf{B} = [-B_0 \sin(2\pi y), B_0 \sin(4\pi x)]$
with $B_0 = 1/\sqrt{4\pi}$. The stabilization diffusion is set to $2.5 \cdot 10^{-5}$ at a resolution of 1000 cells per side length and is decreased linearly together with the cell size refinement. The scaling is performed acoustically by treating of constant CFL number of 0.1.

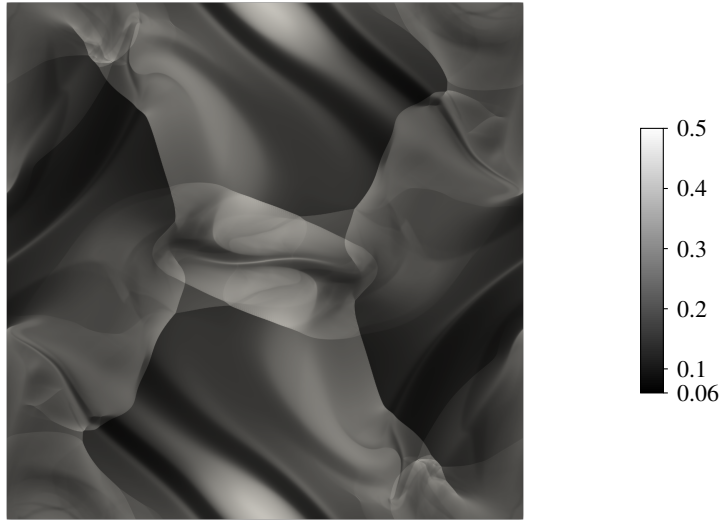
\begin{figure}[htb]
    \centering
    
    % --- Left side: Image ---
    \begin{minipage}[c]{0.45\linewidth}
        \includegraphics[width=\linewidth]{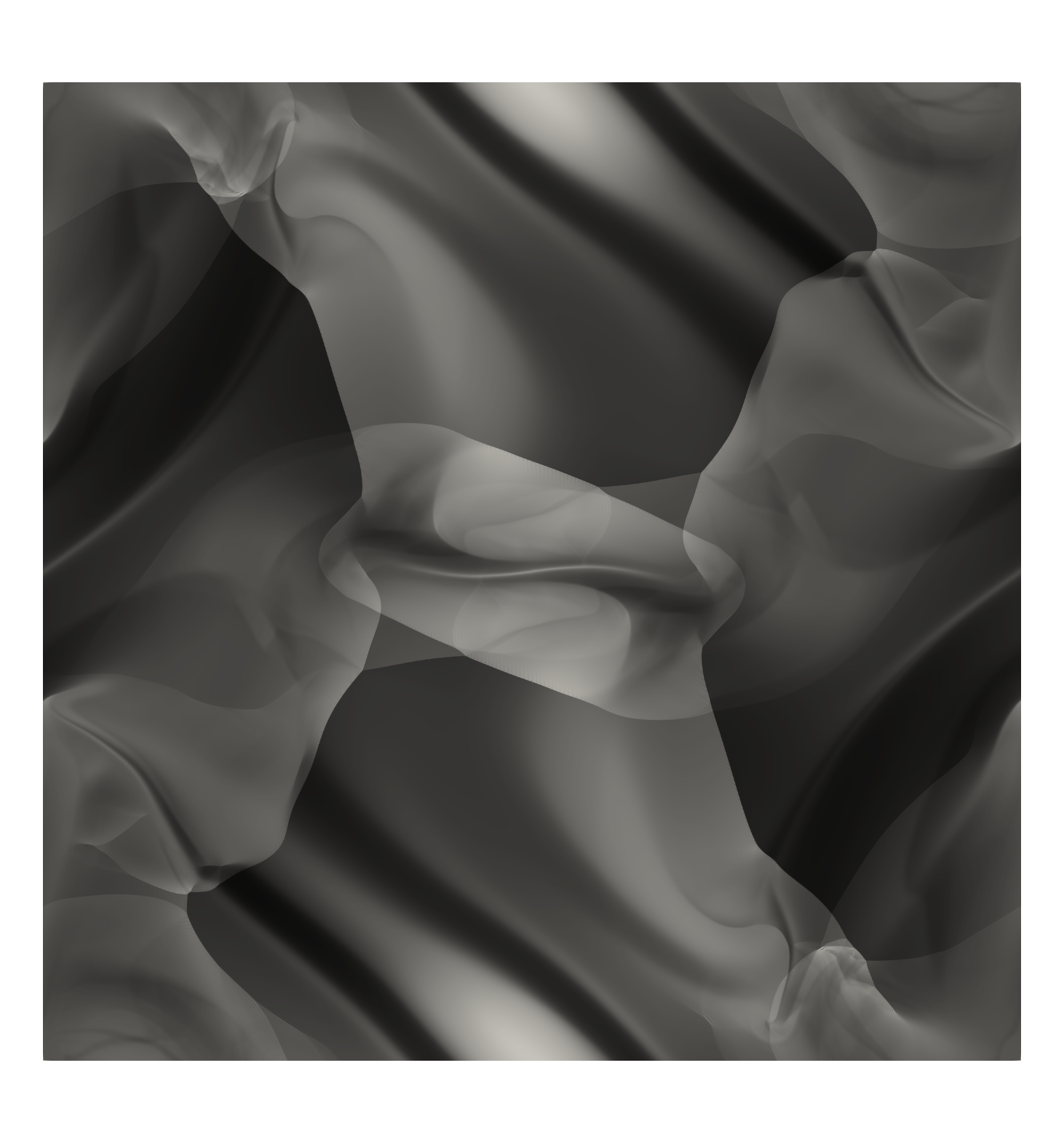}
    \end{minipage}%
    \hspace{0.5cm} 
    % --- Right side: Colorbar ---
    \begin{minipage}[c]{0.15\linewidth} 
        \centering
        \def\minval{0.06}
        \def\maxval{0.5}
        
        \begin{tikzpicture}[y=3.5cm, x=0.3cm] 
            % Draw the black-to-white gradient
            \shade[bottom color=black, top color=white, draw=black] (0,0) rectangle (1, 1);
            
            % Loop through specific values to add ticks and labels
            \foreach \val in {0.06, 0.1, 0.2, 0.3, 0.4, 0.5} {
                % Calculate exact proportional position from 0.0 to 1.0
                \pgfmathsetmacro{\ypos}{(\val - \minval) / (\maxval - \minval)}
                
                % Draw small tick mark
                \draw (1, \ypos) -- (1.2, \ypos);
                % Add the label text
                \node[right, font=\footnotesize] at (1.2, \ypos) {\val};
            }
        \end{tikzpicture}
    \end{minipage}

    \caption{Density distribution in the adiabatic Orszag-Tang vortex benchmark by the resolution of 4000 cells in the side length taken at $t=0.5$, can be compared to \cite{Stone2008} and \cite{Gaburov2011}.}
    \label{fig:compressible_orszag}
\end{figure}

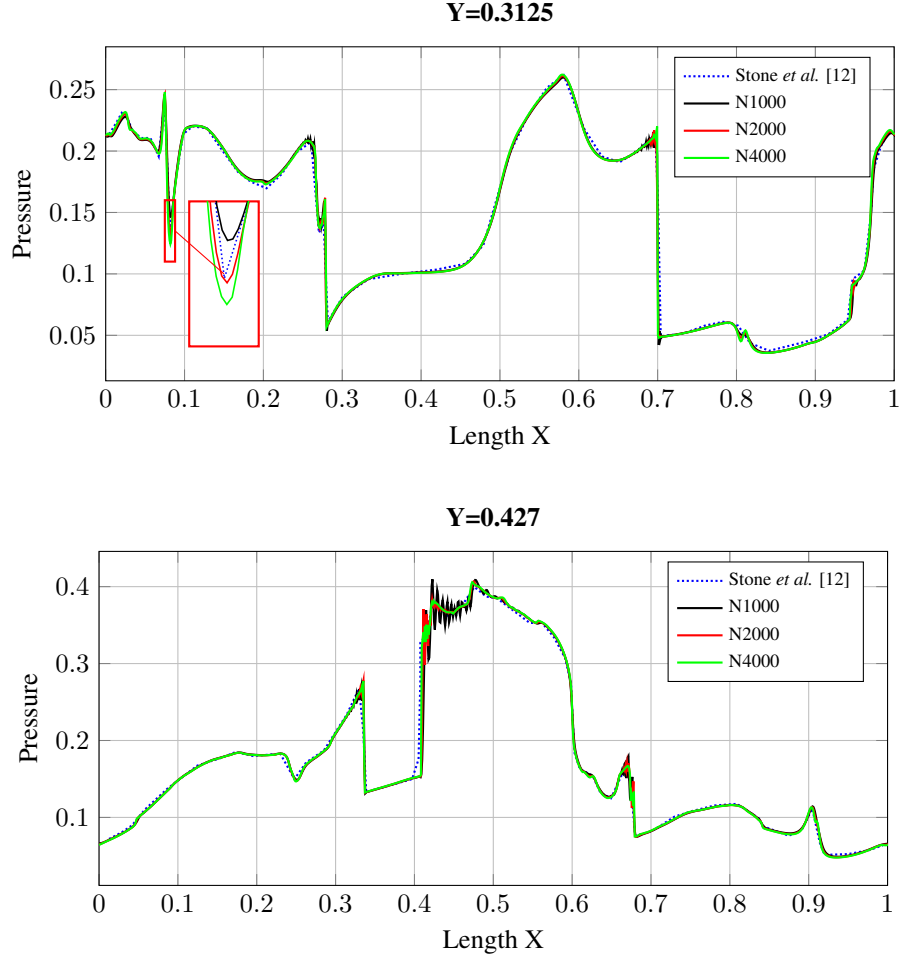
\begin{figure}[htb]
    \centering

    \begin{tikzpicture}
    % ==========================================
    % 1. MAIN PLOT
    % ==========================================
    \begin{axis}[
        name=mainplot,
        width=12cm,
        height=6cm,
        title={\textbf{Y=0.3125}},
        xlabel={Length X},
        ylabel={Pressure},
        xmin = 0,
        xmax = 1,
        legend pos=north east,
        grid=major,
        legend cell align={left},
        ytick distance=0.05,
        yticklabel style={/pgf/number format/fixed},
        legend style={font=\scriptsize}
    ]
        % Reference Data 
        \addplot[mark=none, thick, densely dotted, blue] table [x index=0, y index=1] {ideal_MHD_orszag_tang/p_y03125_ref.csv};
        \addlegendentry{Stone \textit{et al.} \cite{Stone2008}}
        
        % Simulated Data 
        \addplot[mark=none, thick, black] table [x=arc_length, y=Pressure] {ideal_MHD_orszag_tang/N1000_y03125_D2Q5_eta0_f0_5.csv};
        \addlegendentry{N1000}

        \addplot[mark=none, thick, red] table [x=arc_length, y=Pressure] {ideal_MHD_orszag_tang/N2000_y03125_D2Q5_eta0_f0_25.csv};
        \addlegendentry{N2000}

        \addplot[mark=none, thick, green] table [x=arc_length, y=Pressure] {ideal_MHD_orszag_tang/N4000_y03125_D2Q5_eta0_f0_125.csv};
        \addlegendentry{N4000}

        % --- Target Box Formatting ---
        \draw[red, thick] (axis cs:0.075, 0.11) rectangle (axis cs:0.088, 0.16);
        
        \coordinate (insetPosition) at (axis cs:0.15, 0.1); 
        \coordinate (boxEdge) at (axis cs:0.088, 0.135);
    \end{axis}

    % ==========================================
    % 2. INSET PLOT 
    % ==========================================
   \begin{axis}[
        at={(insetPosition)},
        anchor=center,
        width=2.5cm,  
        height=3.5cm, 
        xmin=0.075,   
        xmax=0.088,
        ymin=0.11,    
        ymax=0.16,
        axis background/.style={fill=white}, 
        axis line style={red, thick},        
        xtick=\empty, 
        ytick=\empty,
        enlargelimits=false
    ]
        \addplot[mark=none, semithick, densely dotted, blue] table [x index=0, y index=1] {ideal_MHD_orszag_tang/p_y03125_ref.csv};
        
        \addplot[mark=none, semithick, black] table [x=arc_length, y=Pressure] {ideal_MHD_orszag_tang/N1000_y03125_D2Q5_eta0_f0_5.csv};

        \addplot[mark=none, semithick, red] table [x=arc_length, y=Pressure] {ideal_MHD_orszag_tang/N2000_y03125_D2Q5_eta0_f0_25.csv};

        \addplot[mark=none, semithick, green] table [x=arc_length, y=Pressure] {ideal_MHD_orszag_tang/N4000_y03125_D2Q5_eta0_f0_125.csv};
    \end{axis}
    
    % ==========================================
    % 3. CONNECTING LINE
    % ==========================================
    \draw[red] (boxEdge) -- (insetPosition);

    \end{tikzpicture}

    \vspace{0.5cm}

    \begin{tikzpicture}
        \begin{axis}[
            width=12cm,
            height=6cm,
            title={\textbf{Y=0.427}},
            xlabel={Length X},
            ylabel={Pressure},
            xmin = 0,
            xmax = 1,
            legend pos=north east,
            grid=major,
            legend cell align={left},
            legend style={font=\scriptsize}
        ]
            % Reference Data
            \addplot[mark=none, thick, densely dotted, blue] table [x index=0, y index=1] {ideal_MHD_orszag_tang/p_y0427_ref.csv};
            \addlegendentry{Stone \textit{et al.} \cite{Stone2008}}
            
            % Simulated Data
            \addplot[mark=none, thick, black] table [x=arc_length, y=Pressure] {ideal_MHD_orszag_tang/N1000_y0427_D2Q5_eta0_f0_5.csv};
            \addlegendentry{N1000}

            \addplot[mark=none, thick, red] table [x=arc_length, y=Pressure] {ideal_MHD_orszag_tang/N2000_y0427_D2Q5_eta0_f0_25.csv};
            \addlegendentry{N2000}

            \addplot[mark=none, thick, green] table [x=arc_length, y=Pressure] {ideal_MHD_orszag_tang/N4000_y0427_D2Q5_eta0_f0_125.csv};
            \addlegendentry{N4000}
        \end{axis}
    \end{tikzpicture}
    
    \caption{Validation of the gas pressure values along the horizontal lines at the heights of $y=0.3125$ and $y=0.427$ in the adiabatic Orszag-Tang vortex benchmark taken at $t=0.5$.}
    \label{fig:compressible_orszag_pressure}
\end{figure}

In this compressible regime, the interacting vortices rapidly evolve to produce a system of complex shock waves and strong local current sheets. This configuration serves as a rigorous benchmark for ideal MHD solvers, specifically testing their capacity to sharply resolve fine-scale structures, magnetic reconnection events, and discontinuities. The resulting density distribution is shown in Figure~\ref{fig:compressible_orszag} and the pressure values are provided in Figure~\ref{fig:compressible_orszag_pressure}.

The results show good agreement with the reference values, whereby small oscillations occur at the local shocks at the coarsest resolution due to the chosen stabilization diffusion.
At finer cell sizes, these oscillations disappear, whereby attempts to decrease the stabilization with a higher factor than the cell size refinement factor lead to the divergence of the simulations.
In the upper pressure plot in the zoomed snippet of the curve, it can be noted that the resolutions of 2000 and 4000 show lower numerical diffusion than the reference simulation.

\section{Efficient implementation}\label{sec:performance}

The unique amenability of classical LBMs to efficient parallel execution directly translates to our present approach.
Realizing this potential on diverse target hardware requires an implementation capable of transparently abstracting platform-specific memory and execution hierarchies.
The present approach is implemented within the open-source multi-physics framework OpenLB~\cite{Krause2021b}, which utilizes a hardware-agnostic C++ abstraction layer to achieve performance portability across SIMD CPUs and diverse GPU architectures~\cite{Kummerlaender2023,Kummerlaender2026c}.
By expressing models against abstract cell-level concepts, this design enables compile-time binding to platform-specific data structures.

To bridge the gap between abstract mathematical implementation in C++ and high hardware efficiency, OpenLB employs an automatic optimization pipeline based on \emph{Common Subexpression Elimination} (CSE).
The framework repurposes its differentiable architecture to extract the underlying mathematical expression tree~\cite{Kummerlaender2026c} which is then automatically processed in the SymPy computer algebra system~\cite{Meurer2017} in order to minimize the number of arithmetic operations.

\begin{figure}[p]
    \centering
    \includegraphics[width=0.95\linewidth]{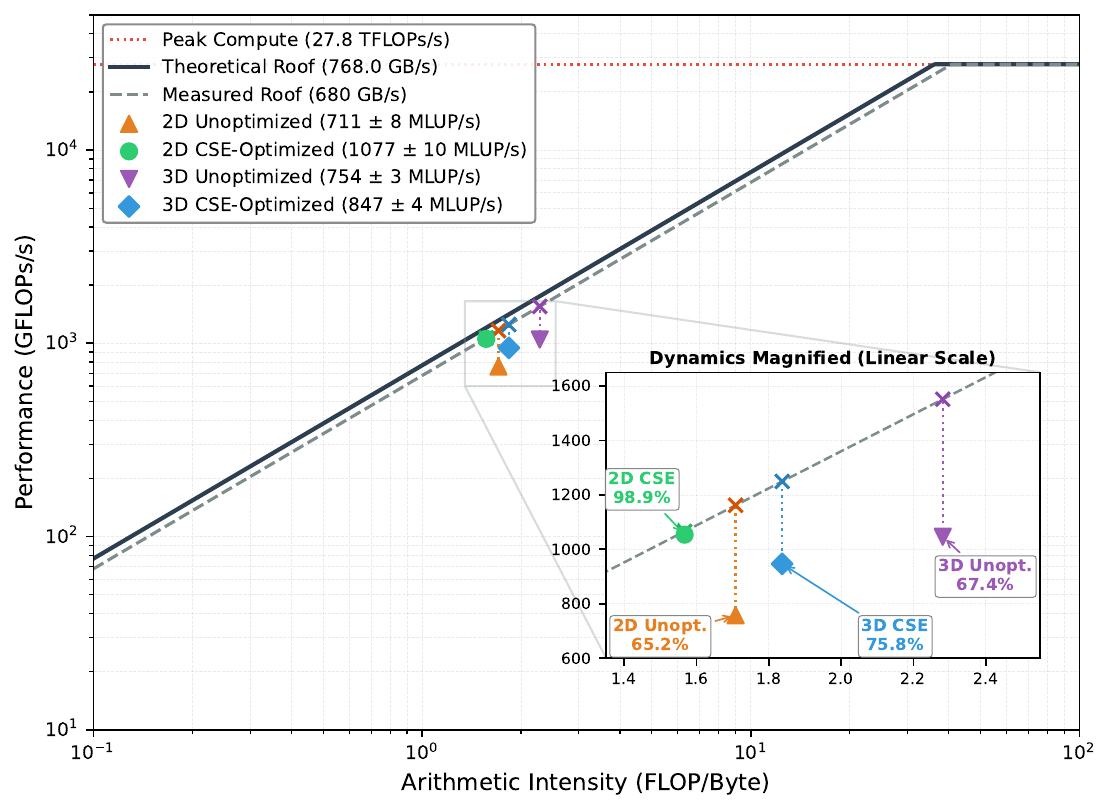}
    \caption{Roofline analysis of both MHD dynamics on NVIDIA A5000 GPU.}
    \label{fig:roofline_analysis}
    \vspace{1.5em}
    \captionof{table}{Performance and roofline analysis of the validated 2D and 3D homogenized MHD collision kernels on a NVIDIA RTX A5000 GPU. Compared to measured peak single-precision memory bandwidth 680\,GB/s (BabelStream)}
    \medskip
    \label{tab:roofline_analysis}
    \begin{tabular}{llllll}
        \toprule
        & \multicolumn{2}{c}{\textbf{2D}} & & \multicolumn{2}{c}{\textbf{3D Homogenized}} \\
        \cmidrule{2-3} \cmidrule{5-6}
        \textbf{Metric} & \textbf{Unoptimized} & \textbf{Optimized} & & \textbf{Unoptimized} & \textbf{Optimized} \\
        \midrule
        Arithmetic Complexity (FLOPs) & 1,066 & 978 & & 1,387 & 1,117 \\
        Operation Reduction (\%)           & --- & 8.26\% & & --- & 19.47\% \\
        Memory Bandwidth (Bytes/cell)      & 624 & 624 & & 608 & 608 \\
        Arithmetic Intensity (FLOP/Byte)   & 1.708 & 1.567 & & 2.281 & 1.837 \\
        \midrule
        Throughput (MLUPs)                & $711 \pm 8$ & $1,077 \pm 10$ & & $754 \pm 3$ & $847 \pm 4$ \\
        Performance (GFLOPs)              & 757.5 & 1,053.7 & & 1046.2 & 946.4 \\
        Roofline Efficiency (\%)           & 65.2\% & 98.9\% & & 67.4\% & 75.8\% \\
        \bottomrule
    \end{tabular}
\end{figure}

Table~\ref{tab:roofline_analysis} demonstrates the impact of this step on an exemplary NVIDIA RTX A5000 GPU.
For the 2D configuration, the CSE pipeline reduces the arithmetic complexity from 1,066 to 978 FLOPs per cell (an 8.26\% reduction), shifting the arithmetic intensity from 1.708 to 1.567 FLOP/Byte at a constant memory bandwidth requirement of 624 Bytes/cell.
Consequently, while the unoptimized 2D kernel achieves 65.2\% of the maximum roofline performance ($711 \pm 8$ MLUP/s), the optimized kernel peaks at an ideal 98.9\% hardware efficiency relative to the roofline peak, processing at $1077 \pm 10$ million lattice updates per second (MLUP/s).
Similarly, the 3D homogenized MHD model sees a 19.47\% reduction in operations (1,387 down to 1,117 FLOPs), reducing the arithmetic intensity from 2.281 to 1.837 FLOP/Byte.
Even when resolving the complex, moving porosity fields representing the solid asteroid geometry at 608 Bytes/cell, the CSE optimized 3D model scales throughput from $754 \pm 3$ to $847 \pm 4$ MLUP/s, achieving a roofline efficiency of 75.8\%.

\section{Solar wind interaction with a magnetized asteroid}\label{sec:asteroid}

To demonstrate the coupled multiphysics capabilities of the present method, we simulate the interaction between a highly magnetized early solar wind and a rotating metallic asteroid modeled after 16 Psyche~\cite{Shepard2021,Nasa2025}.
Such solar wind interactions with planetary bodies are of significant interest in computational astrophysics, traditionally addressed via global MHD or adaptive hybrid kinetic schemes~\cite{Mueller2011, Mejnertsen2018}.

This specific scenario was selected as a challenging computational stress test. Early stellar environments display extreme plasma conditions characterized by high-Mach coronal mass ejections and enhanced magnetic flux densities.
These highly energetic plasma flows drive discontinuities and complex spatial features~\cite{Exner2018, Koenders2015}, testing the numerical stability of our approach.

\begin{figure}[htbp]
    \centering
    \includegraphics[width=0.7\linewidth]{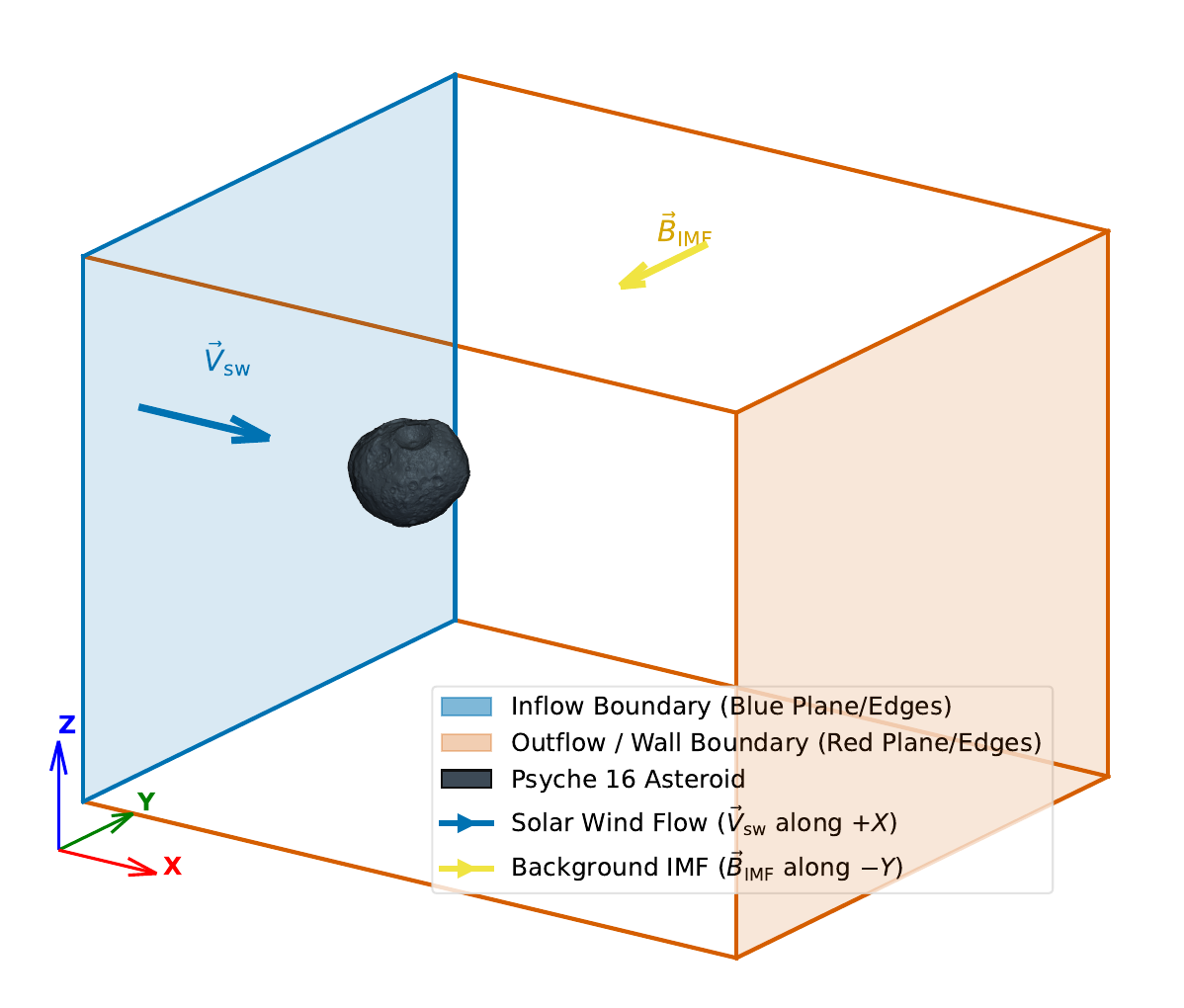}
    \caption{Schematic overview of the 3D computational domain for simulating the interaction of an early solar wind plasma with a magnetized, rotating asteroid. The blue face indicates the supersonic inflow plane ($X_{\text{min}}$), orange faces denote the open boundaries ($X_{\text{max}}$, $Y_{\text{min/max}}$, $Z_{\text{min/max}}$), and the solid central structure represents the homogenized discretization of the asteroid body.}
    \label{fig:psychedomain}
\end{figure}

By embedding a rotating, strongly magnetized solid body within a high-Mach compressible flow using our established homogenized fluid-structure interaction (FSI) approach~\cite{Krause2017,Kummerlaender2026b,Kummerlaender2026a}, the setup encompasses moving boundary conditions, localized dipole fields, and moving shock fronts simultaneously.
The basic setup is illustrated by the schematic overview in Figure~\ref{fig:psychedomain}.

\begin{figure}[p]
    \centering
    \begin{subfigure}{\linewidth}
        \centering
        \includegraphics[width=\linewidth]{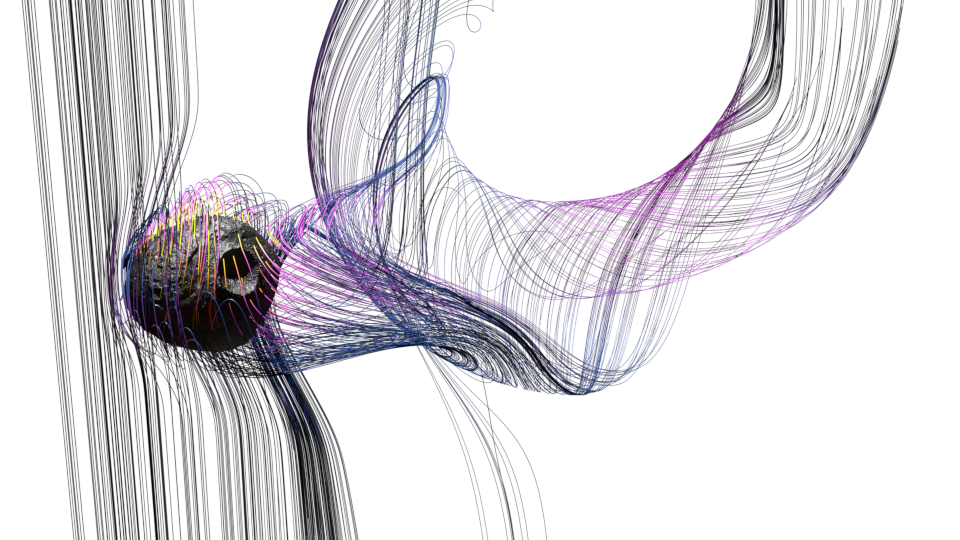}
        \caption{Magnetic field line streamlines showing the complex interaction with internal crustal remanent fields.}
        \label{fig:asteroid_magnetic}
    \end{subfigure}
    \medskip
    \begin{subfigure}{\linewidth}
        \centering
        \includegraphics[width=\linewidth]{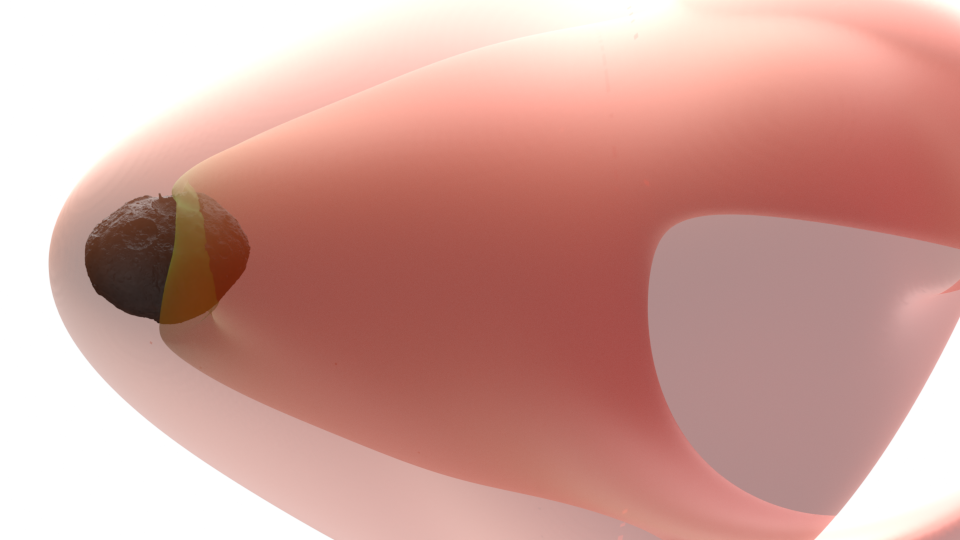}
        \caption{Normalized plasma density contours capturing the bow shock and wake structures.}
        \label{fig:asteroid_density}
    \end{subfigure}
    \caption{Illustrative visualization and rendering~\cite{Marin2025} of the early solar wind interaction with a magnetized, tumbling asteroid modeled after 16 Psyche~\cite{Nasa2025}. The supersonic (Mach 3.5) plasma inflow interacts with the internal crustal remanent fields to produce: (a) a distinct, high-density bow shock and upstream compression region where the frozen-in interplanetary magnetic field (IMF) lines pile up (b) a magnetosheath of twisted flux ropes deflected around the magnetic obstacle (c) an asymmetric, turbulent downstream magnetotail and low-density wake cavity showing magnetic reconnection events.}
    \label{fig:asteroid_combined_showcase}
\end{figure}

To mimic complex crustal remanent features on the metallic asteroid, an analytical ensemble of four independent, shifted, and smoothed magnetic dipoles is embedded inside the reference lattice onto which the Psyche STL geometry~\cite{Nasa2025} is voxelized.
This arrangement scales to a local peak surface field intensity of $225\,\text{nT}$, inducing localized topological stress.
Furthermore, the body is rotated around a normalized arbitrary tilt vector ($\vec{r} = [1, 0, 1]/\sqrt{2}$) with an artificially accelerated angular frequency.
This tumbling behavior continuously twists the frozen-in \emph{interplanetary magnetic field} (IMF) lines (initially oriented along the $-Y$ direction at $5\,\text{nT}$), forcing local magnetic reconnection events, dynamic bow shock structures, and localized Alfvén transients.
The resulting complex plasma topologies and field line structures for a 100 million cell simulation executed on a dual NVIDIA A5000 system are visualized in Figure~\ref{fig:asteroid_combined_showcase}

The inflow boundary ($X_{\text{min}}$) is formulated via an equilibrium projection operator to enforce the continuous, supersonic, and moderately compressible early solar wind fluid state defined in Table~\ref{tab:psyche_parameters}.
The outflow and lateral boundaries ($X_{\text{max}}$, $Y_{\text{min/max}}$, $Z_{\text{min/max}}$) are governed by a directed magnetohydrodynamic zero-gradient boundary condition to allow the plasma flow and transient features to exit the domain without unphysical reflections.

Due to the translation of the porosity-based HLBM approach to MHD, OpenLB's established FSI module~\cite{Kummerlaender2026b,Kummerlaender2026a} can be directly utilized both to efficiently project the rotating geometry and its remnant magnetic field into the fluid lattice, and to integrate various quantities of interest at every timestep. For the present setup we obtained per-timestep surface integrals of:
\begin{itemize}
    \item Net magnetic flux ($\text{Wb}$) passing through the asteroid.
    \item Forces ($\text{N}$), comprising both aerodynamic and Maxwell electromagnetic forces.
    \item Torques ($\text{N}\cdot\text{m}$) acting on the rotating asteroid.
    \item Electromagnetic work ($\text{J}$) transferred to/from the solar wind.
\end{itemize}    

\begin{table}
\centering
\caption{Simulation parameters for the 16 Psyche early solar wind interaction showcase.}
\label{tab:psyche_parameters}
\begin{tabular}{@{}ll@{}}
\toprule
\textbf{Parameter} & \textbf{Value} \\
\midrule
Asteroid Diameter & $220$ km \\
Solar Wind Velocity & $400$ km/s \\
Solar Wind Density & $8.35 \times 10^{-21}$ kg/m$^3$ \\
Interplanetary Magnetic Field (IMF) & $5$ nT \\
Maximum Crustal Magnetic Field & $250$ nT \\
Simulation Mach Number & $3.5$ \\
CFL & $0.2$ \\
Lattice Relaxation Time & $0.51$ \\
Cell count & $100 \times 10^{6}$ \\
Lattice velocities set & D3Q7 \\
Collision operator & BGK \\
Source scheme & Direct \\
\bottomrule
\end{tabular}
\end{table}

While this configuration is designed as an algorithmic capability and high-performance computing showcase rather than a phenomenologically validated planetary study, it together with the previous rigorous benchmarks demonstrates the viability of our strictly local scheme for such applications.

\section{Conclusion}

In this work, we introduced a novel, fully local, partial differential equation (PDE) agnostic Lattice Boltzmann Method (LBM) framework for solving generic systems of conservation laws.
We demonstrated the capability of this approach by applying it to highly complex multiphysics problems, specifically compressible ideal and incompressible resistive magnetohydrodynamics (MHD).
By expanding the macroscopic state vector to transport spatial gradients and viscous stresses as independent variables, the scheme natively recovers required derivatives while completely avoiding non-local finite-difference operations.
In the context of MHD, this local evolution seamlessly supports the enforcement of the solenoidal constraint on the magnetic field.
Validation across established benchmarks, including the Brio-Wu shock tube, the MHD rotor, and the Orszag-Tang vortex, confirms the framework's high accuracy, stability, and robust discontinuity resolution.
Furthermore, implementation within the OpenLB platform, augmented by automated common subexpression elimination, translates this localized algorithmic structure into exceptional computational efficiency, achieving up to 98.9\% of the hardware roofline limit on a modern GPU architecture.
Finally, our 3D astrophysical application of early solar wind interacting with a tumbling, magnetized asteroid demonstrates the scheme's capacity for complex, large-scale fluid-structure interactions, establishing this PDE-agnostic approach as a highly scalable and robust tool for advanced multiphysics simulations.

\section*{Artificial intelligence statement}

Google Gemini was utilized by the authors for code generation and text formatting assistance during the preparation of this work.
The final manuscript was reviewed, edited, and approved by all authors, who maintain full accountability for the content.

\newpage

\bibliographystyle{unsrt}  
\bibliography{bibliography}

@article{Anandan2024,
  title={On Lattice Boltzmann Methods based on vector-kinetic models for hyperbolic partial differential equations},
  author={Anandan, Megala and Raghurama Rao, S. V.},
  journal={Computers \& Fluids},
  volume={280},
  pages={106348},
  year={2024},
  publisher={Elsevier}
}

@article{Brio1988,
title = {An upwind differencing scheme for the equations of ideal magnetohydrodynamics},
journal = {Journal of Computational Physics},
volume = {75},
number = {2},
pages = {400-422},
year = {1988},
issn = {0021-9991},
doi = {https://doi.org/10.1016/0021-9991(88)90120-9},
url = {https://www.sciencedirect.com/science/article/pii/0021999188901209},
author = {M Brio and C.C Wu},
}

@article{Chai2013,
  title = {Lattice Boltzmann model for the convection-diffusion equation},
  author = {Chai, Zhenhua and Zhao, T. S.},
  journal = {Phys. Rev. E},
  volume = {87},
  issue = {6},
  pages = {063309},
  numpages = {15},
  year = {2013},
  month = {Jun},
  publisher = {American Physical Society},
  doi = {10.1103/PhysRevE.87.063309},
  url = {https://link.aps.org/doi/10.1103/PhysRevE.87.063309}
}

@article{Dellar2013,
title = {Lattice Boltzmann magnetohydrodynamics with current-dependent resistivity},
journal = {Journal of Computational Physics},
volume = {237},
pages = {115-131},
year = {2013},
issn = {0021-9991},
doi = {https://doi.org/10.1016/j.jcp.2012.11.021},
url = {https://www.sciencedirect.com/science/article/pii/S0021999112007012},
author = {Paul J. Dellar},
}

@article{Gaburov2011,
  title={Astrophysical weighted particle magnetohydrodynamics},
  author={Gaburov, Evghenii and Nitadori, Keigo},
  journal={Monthly Notices of the Royal Astronomical Society},
  volume={414},
  number={1},
  pages={129--154},
  year={2011},
  publisher={Oxford University Press}
}

@article{Grad1949,
author = {Grad, Harold},
title = {On the kinetic theory of rarefied gases},
journal = {Communications on Pure and Applied Mathematics},
volume = {2},
number = {4},
pages = {331-407},
doi = {https://doi.org/10.1002/cpa.3160020403},
url = {https://onlinelibrary.wiley.com/doi/abs/10.1002/cpa.3160020403},
eprint = {https://onlinelibrary.wiley.com/doi/pdf/10.1002/cpa.3160020403},
year = {1949}
}

@article{Huang2024,
  title={Modeling of nonequilibrium effects in a compressible plasma based on the lattice Boltzmann method},
  author={Huang, Haoyu and Jin, Ke and Li, Kai and Zheng, Xiaojing},
  journal={Physics of Plasmas},
  volume={31},
  number={9},
  pages={093902},
  year={2024},
  publisher={AIP Publishing}
}

@Article{Krause2017,
  author   = {Mathias J. Krause and Fabian Klemens and Thomas Henn and Robin Trunk and Hermann Nirschl},
  journal  = {Particuology},
  title    = {Particle flow simulations with homogenised lattice Boltzmann methods},
  year     = {2017},
  issn     = {1674-2001},
  pages    = {1-13},
  volume   = {34},
  doi      = {https://doi.org/10.1016/j.partic.2016.11.001},
  url      = {https://www.sciencedirect.com/science/article/pii/S167420011730041X},
}

@Book{Krueger2016,
  author    = {Timm Krueger and Halim Kusumaatmaja and Alexandr Kuzmin and Orest Shardt and Goncalo Silva and Viggen, {Erlend Magnus}},
  publisher = {Springer},
  title     = {The Lattice Boltzmann Method: Principles and Practice},
  year      = {2016},
  isbn      = {978-3-319-44647-9},
  series    = {Graduate Texts in Physics},
  language  = {English},
}

@article{Kummerlaender2025,
  author    = {Kummerl{\"a}nder, Adrian and Bukreev, Fedor and Teutscher, Dennis and Dorn, M{\'a}rcio and Krause, Mathias J.},
  title     = {Optimization of Single Node Load Balancing for {Lattice} {Boltzmann} {Method} on Heterogeneous High Performance Computers},
  journal   = {Journal of Parallel and Distributed Computing},
  year      = {2025},
  volume    = {206},
  doi       = {10.1016/j.jpdc.2025.105169}
}

@article{Kummerlaender2026a,
  author    = {Kummerl{\"a}nder, Adrian and Tur, B. and Haase, M. and Bukreev, Fedor and D{\"o}llinger, M. and Krause, Mathias J. and Kniesburges, S.},
  title     = {Efficient fluid structure interaction simulation of vocal fold oscillations using a homogenized {Lattice} {Boltzmann} {Method}},
  journal   = {Computer Methods in Applied Mechanics and Engineering},
  year      = {2026},
  volume    = {457},
  pages     = {119009},
  doi       = {10.1016/j.cma.2026.119009}
}

@article{Kummerlaender2026b,
  author    = {Kummerl{\"a}nder, Adrian and Ito, Shota and Schecher, Maximilian and Dapelo, Davide and Simonis, Stephan and Krause, Mathias J. and Bukreev, Fedor},
  title     = {Efficient Wall-Modelled Large Eddy Simulation of Rotors using Homogenized {Lattice} {Boltzmann} {Methods}},
  journal   = {International Journal of Numerical Methods for Heat \& Fluid Flow},
  year      = {2026},
  doi       = {10.1108/HFF-09-2025-0724}
}

@preprint{Kummerlaender2026c,
  author    = {Adrian Kummerl{\"a}nder and Fedor Bukreev and Leonardo Dorneles and Marcio Dorn and Shota Ito and Mathias J. Krause},
  title     = {A Hardware Abstraction for Exascale Lattice Boltzmann Simulations: Rotor-Resolved Wind Farms on Aurora and LUMI},
  year      = {2026}
}

@article{Kummerlaender2023,
  author    = {Kummerl{\"a}nder, Adrian and Dorn, M{\'a}rcio and Frank, Martin and Krause, Mathias J.},
  title     = {Implicit Propagation of Directly Addressed Grids in {Lattice} {Boltzmann} {Methods}},
  journal   = {Concurrency and Computation: Practice and Experience},
  year      = {2023},
  volume    = {35},
  number    = {8},
  pages     = {e7509},
  doi       = {10.1002/cpe.7509}
}

@article{Liu2017,
  title={A Unified Gas Kinetic Scheme for Continuum and Rarefied Flows V: Multiscale and Multi-Component Plasma Transport},
  author={Liu, Chang and Xu, Kun},
  journal={Communications in Computational Physics},
  volume={22},
  number={5},
  pages={1175--1223},
  year={2017},
  publisher={Global Science Press}
}

@article{Londrillo2000,
  title={High-order upwind schemes for multidimensional magnetohydrodynamics},
  author={Londrillo, Pasquale and Del Zanna, Luca},
  journal={The Astrophysical Journal},
  volume={530},
  number={1},
  pages={508--524},
  year={2000},
  publisher={IOP Publishing}
}

@article{Coveney2002,
    author = {Coveney, P. V. and Succi, S. and d'Humières, Dominique and Ginzburg, Irina and Krafczyk, Manfred and Lallemand, Pierre and Luo, Li-Shi},
    title = {Multiple–relaxation–time lattice Boltzmann models in three dimensions},
    journal = {Philosophical Transactions of the Royal Society A: Mathematical, Physical and Engineering Sciences},
    volume = {360},
    number = {1792},
    pages = {437-451},
    year = {2002},
    month = {03},
    issn = {1364-503X},
    doi = {10.1098/rsta.2001.0955},
    url = {https://doi.org/10.1098/rsta.2001.0955},
    eprint = {https://royalsocietypublishing.org/rsta/article-pdf/360/1792/437/323208/rsta.2001.0955.pdf},
}

@article{Krause2021a,
 author = {Krause, Mathias J. and Kummerl{\"a}nder, Adrian and Avis, Samuel J. and Kusumaatmaja, Halim and Dapelo, Davide and Klemens, Fabian and Gaedtke, Maximilian and Hafen, Nicolas and Mink, Albert and Trunk, Robin and Marquardt, Jan E. and Maier, Marie Luise and Haussmann, Marc and Simonis, Stephan},
 year = {2021},
 title = {OpenLB---Open source lattice Boltzmann code},
 pages = {258--288},
 volume = {81},
 issn = {0898-1221},
 journal = {Computers {\&} Mathematics with Applications},
 doi = {10.1016/J.CAMWA.2020.04.033},
}

@article{Powell1999,
title = {A Solution-Adaptive Upwind Scheme for Ideal Magnetohydrodynamics},
journal = {Journal of Computational Physics},
volume = {154},
number = {2},
pages = {284-309},
year = {1999},
issn = {0021-9991},
doi = {https://doi.org/10.1006/jcph.1999.6299},
url = {https://www.sciencedirect.com/science/article/pii/S002199919996299X},
author = {Kenneth G. Powell and Philip L. Roe and Timur J. Linde and Tamas I. Gombosi and Darren L. {De Zeeuw}},
}

@article{Struchtrup2003,
    author = {Struchtrup, Henning and Torrilhon, Manuel},
    title = {Regularization of Grad’s 13 moment equations: Derivation and linear analysis},
    journal = {Physics of Fluids},
    volume = {15},
    number = {9},
    pages = {2668-2680},
    year = {2003},
    month = {09},
    issn = {1070-6631},
    doi = {10.1063/1.1597472},
    url = {https://doi.org/10.1063/1.1597472},
    eprint = {https://pubs.aip.org/aip/pof/article-pdf/15/9/2668/19040326/2668_1_online.pdf},
}

@article{Latt2006,
author = {Latt, Jonas and Chopard, Bastien},
title = {Lattice Boltzmann method with regularized pre-collision distribution functions},
year = {2006},
issue_date = {9 September 2006},
publisher = {Elsevier Science Publishers B. V.},
address = {NLD},
volume = {72},
number = {2–6},
issn = {0378-4754},
url = {https://doi.org/10.1016/j.matcom.2006.05.017},
doi = {10.1016/j.matcom.2006.05.017},
journal = {Math. Comput. Simul.},
month = sep,
pages = {165–168},
numpages = {4},
}

@article{Stone2008,
doi = {10.1086/588755},
url = {https://doi.org/10.1086/588755},
year = {2008},
month = {sep},
publisher = {},
volume = {178},
number = {1},
pages = {137},
author = {Stone, James M. and Gardiner, Thomas A. and Teuben, Peter and Hawley, John F. and Simon, Jacob B.},
title = {Athena: A New Code for Astrophysical MHD},
journal = {The Astrophysical Journal Supplement Series},
}

@article{Ginzburg2018,
  author  = {Ginzburg, I. and d'Humières, D. and Kuzmin, A.},
  title   = {Two-Relaxation-Time Lattice Boltzmann Scheme: About Parametrization, Velocity, Pressure and Mixed Boundary Conditions},
  journal = {Communications in Computational Physics},
  year    = {2018},
  month   = {Mar.},
  volume  = {3},
  number  = {2},
  pages   = {427--478},
  doi     = {10.4208/cicp.2008.v3.p427},
  url     = {https://www.global-sci.com/cicp/article/view/5523}
}

@article{Wissocq2024,
  title={A positive- and bound-preserving vectorial lattice Boltzmann method in two dimensions},
  author={Wissocq, Gauthier and Liu, Yongle and Abgrall, R{\'e}mi},
  journal={arXiv preprint arXiv:2411.15001},
  year={2024}
}

@article{Angot1999,
  title = {A Penalization Method to Take into Account Obstacles in Incompressible Viscous Flows},
  author = {Angot, Philippe and Bruneau, Charles-Henri and Fabrie, Pierre},
  year = 1999,
  month = feb,
  journal = {Numerische Mathematik},
  volume = {81},
  number = {4},
  pages = {497--520},
  issn = {0945-3245},
  doi = {10.1007/s002110050401},
  langid = {english},
}

@article{Bhatnagar1954,
  title = {A model for collision processes in gases. I. Small amplitude processes in charged and neutral one-component systems},
  author = {Bhatnagar, P. L. and Gross, E. P. and Krook, M.},
  journal = {Phys. Rev.},
  volume = {94},
  issue = {3},
  pages = {511--525},
  numpages = {0},
  year = {1954},
  month = {May},
  publisher = {American Physical Society},
  doi = {10.1103/PhysRev.94.511},
  url = {https://link.aps.org/doi/10.1103/PhysRev.94.511}
}

@article{Boolakee2025,
title = {Lattice Boltzmann for linear elastodynamics: Periodic problems and Dirichlet boundary conditions},
journal = {Computer Methods in Applied Mechanics and Engineering},
volume = {433},
pages = {117469},
year = {2025},
issn = {0045-7825},
doi = {https://doi.org/10.1016/j.cma.2024.117469},
url = {https://www.sciencedirect.com/science/article/pii/S0045782524007242},
author = {Oliver Boolakee and Martin Geier and Laura {De Lorenzis}},
}

@article{Dubois2014,
  TITLE = {{Simulation of strong nonlinear waves with vectorial lattice Boltzmann schemes}},
  AUTHOR = {Dubois, Fran{\c c}ois},
  URL = {https://hal.science/hal-00923281},
  NOTE = {12 pages},
  JOURNAL = {{International Journal of Modern Physics C}},
  PUBLISHER = {{World Scientific Publishing}},
  VOLUME = {25},
  NUMBER = {12},
  PAGES = {1441014},
  YEAR = {2014},
  DOI = {10.1142/S0129183114410149},
  HAL_ID = {hal-00923281},
  HAL_VERSION = {v2},
}

@article{Exner2018,
  title = {Coronal Mass Ejection Hits Mercury: {{A}}.{{I}}.{{K}}.{{E}}.{{F}}. Hybrid-Code Results Compared to {{MESSENGER}} Data},
  shorttitle = {Coronal Mass Ejection Hits Mercury},
  author = {Exner, W. and Heyner, D. and Liuzzo, L. and Motschmann, U. and Shiota, D. and Kusano, K. and Shibayama, T.},
  year = 2018,
  month = apr,
  journal = {Planetary and Space Science},
  volume = {153},
  pages = {89--99},
  issn = {0032-0633},
  doi = {10.1016/j.pss.2017.12.016},
  urldate = {2026-05-28},
}

@misc{Guillon2024,
      title={Stability analysis of the vectorial lattice-Boltzmann method}, 
      author={Kévin Guillon and Romane Hélie and Philippe Helluy},
      year={2024},
      eprint={2402.09813},
      archivePrefix={arXiv},
      primaryClass={math.AP},
      url={https://arxiv.org/abs/2402.09813}, 
}

@article{Guo2002,
  title = {Discrete lattice effects on the forcing term in the lattice Boltzmann method},
  author = {Guo, Zhaoli and Zheng, Chuguang and Shi, Baochang},
  journal = {Phys. Rev. E},
  volume = {65},
  issue = {4},
  pages = {046308},
  numpages = {6},
  year = {2002},
  month = {Apr},
  publisher = {American Physical Society},
  doi = {10.1103/PhysRevE.65.046308},
  url = {https://link.aps.org/doi/10.1103/PhysRevE.65.046308}
}

@article{Koenders2015,
  title = {Dynamical Features and Spatial Structures of the Plasma Interaction Region of {{67P}}/{{Churyumov}}--{{Gerasimenko}} and the Solar Wind},
  author = {Koenders, C. and Glassmeier, K. -H. and Richter, I. and Ranocha, H. and Motschmann, U.},
  year = 2015,
  month = jan,
  journal = {Planetary and Space Science},
  volume = {105},
  pages = {101--116},
  issn = {0032-0633},
  doi = {10.1016/j.pss.2014.11.014},
  urldate = {2026-05-27},
}

@article{Krause2021b,
 author = {Krause, Mathias J. and Kummerl{\"a}nder, Adrian and Avis, Samuel J. and Kusumaatmaja, Halim and Dapelo, Davide and Klemens, Fabian and Gaedtke, Maximilian and Hafen, Nicolas and Mink, Albert and Trunk, Robin and Marquardt, Jan E. and Maier, Marie-Luise and Haussmann, Marc and Simonis, Stephan},
 year = {2021},
 title = {OpenLB---Open source lattice Boltzmann code},
 journal = {Computers {\&} Mathematics with Applications},
 doi = {10.1016/j.camwa.2020.04.033},
}

@article{Mejnertsen2018,
  title = {Global {{MHD Simulations}} of the {{Earth}}'s {{Bow Shock Shape}} and {{Motion Under Variable Solar Wind Conditions}}},
  author = {Mejnertsen, L. and Eastwood, J. P. and Hietala, H. and Schwartz, S. J. and Chittenden, J. P.},
  year = 2018,
  month = jan,
  journal = {Journal of Geophysical Research: Space Physics},
  volume = {123},
  number = {1},
  pages = {259--271},
  issn = {2169-9380, 2169-9402},
  doi = {10.1002/2017JA024690},
  urldate = {2026-05-16},
  copyright = {http://creativecommons.org/licenses/by/4.0/},
  langid = {english},
}

@article{Mueller2011,
  title = {A.{{I}}.{{K}}.{{E}}.{{F}}.: {{Adaptive}} Hybrid Model for Space Plasma Simulations},
  shorttitle = {A.{{I}}.{{K}}.{{E}}.{{F}}.},
  author = {M{\"u}ller, Joachim and Simon, Sven and Motschmann, Uwe and Sch{\"u}le, Josef and Glassmeier, Karl-Heinz and Pringle, Gavin J.},
  year = 2011,
  month = apr,
  journal = {Computer Physics Communications},
  volume = {182},
  number = {4},
  pages = {946--966},
  issn = {0010-4655},
  doi = {10.1016/j.cpc.2010.12.033},
  urldate = {2026-05-28},
}

@misc{Nasa2025,
  author       = {{NASA Science}},
  title        = {Asteroid {P}syche 3{D} {M}odel and {G}eometry},
  howpublished = {\url{https://science.nasa.gov/solar-system/asteroids/16-psyche/}},
  year         = {2025},
  note         = {Accessed: 2026-05-28}
}

@article{Marin2025,
title = {SciBlend: Advanced data visualization workflows within Blender},
journal = {Computers \& Graphics},
volume = {130},
pages = {104264},
year = {2025},
issn = {0097-8493},
doi = {https://doi.org/10.1016/j.cag.2025.104264},
url = {https://www.sciencedirect.com/science/article/pii/S0097849325001050},
author = {José Marín and Tiffany M.G. Baptiste and Cristobal Rodero and Steven E. Williams and Steven A. Niederer and Ignacio García-Fernández},
}

@article{Shepard2021,
  title = {Asteroid 16 {{Psyche}}: {{Shape}}, {{Features}}, and {{Global Map}}},
  shorttitle = {Asteroid 16 {{Psyche}}},
  author = {Shepard, Michael K. and De Kleer, Katherine and Cambioni, Saverio and Taylor, Patrick A. and Virkki, Anne K. and {R{\'i}vera-Valentin}, Edgard G. and {Rodriguez Sanchez-Vahamonde}, Carolina and {Fernanda Zambrano-Marin}, Luisa and Magri, Christopher and Dunham, David and Moore, John and Camarca, Maria},
  year = 2021,
  month = aug,
  journal = {The Planetary Science Journal},
  volume = {2},
  number = {4},
  pages = {125},
  issn = {2632-3338},
  doi = {10.3847/PSJ/abfdba},
  urldate = {2026-05-28},
  langid = {english},
}

@article{Strang1968,
author = {Strang, Gilbert},
title = {On the Construction and Comparison of Difference Schemes},
journal = {SIAM Journal on Numerical Analysis},
volume = {5},
number = {3},
pages = {506-517},
year = {1968},
doi = {10.1137/0705041}}

@article{Meurer2017,
     title = {SymPy: symbolic computing in Python},
     author = {Meurer, Aaron and Smith, Christopher P. and Paprocki, Mateusz and \v{C}ert\'{i}k, Ond\v{r}ej and Kirpichev, Sergey B. and Rocklin, Matthew and Kumar, AMiT and Ivanov, Sergiu and Moore, Jason K. and Singh, Sartaj and Rathnayake, Thilina and Vig, Sean and Granger, Brian E. and Muller, Richard P. and Bonazzi, Francesco and Gupta, Harsh and Vats, Shivam and Johansson, Fredrik and Pedregosa, Fabian and Curry, Matthew J. and Terrel, Andy R. and Rou\v{c}ka, \v{S}t\v{e}p\'{a}n and Saboo, Ashutosh and Fernando, Isuru and Kulal, Sumith and Cimrman, Robert and Scopatz, Anthony},
     year = 2017,
     month = jan,
     volume = 3,
     pages = {e103},
     journal = {PeerJ Computer Science},
     issn = {2376-5992},
     url = {https://doi.org/10.7717/peerj-cs.103},
     doi = {10.7717/peerj-cs.103}
    }

\end{document}